\documentclass[preprint2]{aastex}

\newcommand{\aped} {APED}
\newcommand{\arl} {AR~Lac}
\newcommand{\asca} {{\it ASCA}}
\newcommand{\chan} {{\it Chandra}}
\newcommand{\ciao} {CIAO}
\newcommand{\cxc} {CXC}

\newcommand{\dem} {DEM}
\newcommand{\euv} {EUV}
\newcommand{\euve} {{\it EUVE}}
\newcommand{\exosat} {{\it EXOSAT}}
\newcommand{\fwhm} {FWHM}
\newcommand{\heg}  {HEG}
\newcommand{\hetgs} {HETGS}
\newcommand{\hetg} {HETG}
\newcommand{\isis} {ISIS}

\newcommand{\meg}  {MEG}
\newcommand{\obs}  {OID}

\newcommand{\rosat} {{\it ROSAT}}
\newcommand{\sax} {{\it Beppo-SAX}}

\newcounter{ion}
\newcommand{\eli}[2]{\setcounter{ion}{#2}#1{~\sc\roman{ion}}}
\newcommand{\mone}{^{-1}}
\newcommand{\mtwo}{^{-2}}
\newcommand{\mthree}{^{-3}}

\newcommand{\kms}{\mathrm{\,km\,s\mone}}

\newcommand{\rev}[1]{{\bf #1}}    
\renewcommand{\rev}[1]{#1}        
\begin{document}

\title{The Coronae of AR Lac}
\shorttitle{The Coronae of AR Lac}
\shortauthors{Huenemoerder et al.}
\slugcomment{To appear in ApJ }

\author{David P.\ Huenemoerder,  Claude R.\ Canizares}
\affil{Massachusetts Institute of Technology, Cambridge, MA 02139}
\email{dph@space.mit.edu, crc@space.mit.edu}

\author{Jeremy J.\ Drake}
\affil{Harvard-Smithsonian Center for Astrophysics, Cambridge, MA 02138}
\email{jdrake@head-cfa.harvard.edu}

\and

\author{Jorge Sanz-Forcada}
\affil{Osservatorio Astronomico di Palermo
G.S. Vaiana, Piazza del Palramento, 1; Palermo, I-90134, Italy}
\email{jsanz@astropa.unipa.it}

\begin{abstract}
  
  We observed the coronally active eclipsing binary \arl\ with the
  High Energy Transmission Grating on \chan\ for a total of 97 ks,
  spaced over five orbits, at quadratures and conjunctions.
  Contemporaneous and simultaneous EUV spectra and photometry were
  also obtained with the Extreme Ultraviolet Explorer.  Significant
  variability in both X-ray and EUV fluxes were observed, dominated by
  at least one X-ray flare and one EUV flare.  We saw no evidence of
  primary or secondary eclipses, but exposures at these phases were
  short and intrinsic variability compromised detection of any
  geometric modulation.  X-ray flux modulation was largest at high
  temperature, indicative of flare heating of coronal plasma rather
  than changes in emitting volume or global emission measure.
  Analysis of spectral line widths interpreted in terms of Doppler
  broadening suggests that both binary stellar components are active.
  Based on line fluxes obtained from total integrated spectra, we have
  modeled the emission measure and abundance distributions.  The EUV
  spectral line fluxes were particularly useful for constraining the
  parameters of the ``cool'' ($\leq 2\times 10^6$~K) plasma. A strong
  maximum was found in the differential emission measure,
  characterized by two apparent peaks at $\log T = 6.9$ and $7.4$,
  together with a weak but significant cooler maximum near $\log
  T=6.2$, and a moderately strong hot tail from $\log T= 7.6-8.2$.
  Coronal abundances have a broad distribution and show no simple
  correlation with first ionization potential.  While the resulting
  model spectrum generally agrees very well with the observed
  spectrum, there are some significant discrepancies, especially among
  the many Fe L-lines.  Both the emission measure and abundance
  distributions are qualitatively similar to prior determinations from
  other X-ray and ultraviolet spectra, indicating some long-term
  stability in the overall coronal structure.

\end{abstract}


\section{Introduction}\label{sec:intro}

\arl\ (HD~210334, HR~8448) is one of the brightest totally eclipsing
RS~CVn binaries.  Since eclipses can help constrain active region
geometry, it has been a key system for studying the structure of
photospheric spots from visible light modulation, the chromosphere
from emission of magnesium, calcium and hydrogen, the transition
region through ultraviolet emission lines, and the coronae via
emission at extreme ultraviolet (EUV), X-ray, and radio wavelengths.
There is as yet no comprehensive predictive theory that explains in detail
all the aspects of coronal emission based only on fundamental stellar
parameters.  Observational attack is then aimed at 
increasing the quantity and quality of spectral and photometric data to
provide insights and to help constrain the dependence of coronal activity on
stellar evolutionary parameters, other magnetic
activity indicators or as yet unidentified parameters.  With the
High Energy Transmission Grating Spectrometer (\hetgs) on the \chan\
X-Ray Observatory, we are able to greatly improve the quality of X-ray
spectra by resolving a multitude of coronal emission lines due to iron
and the hydrogen-like and helium-like lines of a number of abundant
elements.  These spectra provide both line and continuum fluxes and
their time variability.  The aim of this study is to model the lines
and continuum with the latest atomic data in order to determine the
coronal temperature structure, density and absolute elemental
abundances.

\arl\ is comprised of G- and K-type subgiants in a 1.98 day orbit.
The components are each of slightly greater than one Solar mass
($1.23M_\odot$ and $1.27M_\odot$, respectively) and have radii of
$1.52R_\odot$ and $2.72R_\odot$.  They reach a maximum radial velocity
separation of $230\,\mathrm{km\,s\mone}$ and have rotational
velocities of 39 and 70 $\mathrm{km\,s\mone}$.  At a distance of 42
pc, \arl\ is relatively bright.  \citet{Gehren:99} summarized these and
other fundamental properties of \arl.

\arl\ was detected in X-rays by HEAO 1 at a luminosity of
$\sim10^{31}\,\mathrm{ergs\,s\mone}$ \citep{Walter:80};
\citet{Walter:83} present an early analysis of the X-ray coronae in
which it was inferred that emission arose from both stellar components
in compact and extended structures.  Subsequent X-ray studies were
undertaken by \citet{Ottmann:Schmitt:al:93} (\rosat),
\citet{White:Shafer:al:90} (\exosat), \citet{Rodono:99} (\sax), and
\citet{White:Arnaud:al:94} (\asca).

Observations with \rosat\ and \asca\ detected a deep primary eclipse
and a smaller secondary eclipse
\citep{Ottmann:Schmitt:94,White:Arnaud:al:94}, while \exosat\ and
\euve\ observations could only confirm the primary eclipse due to
flares or instrumental limitations
\citep{White:Shafer:al:90,Christian:Drake:al:96,Brickhouse:Dupree:al:99,Pease:Drake:al:02}.
Some extended chromospheric material was also detected by
\citet{Montes:Fernandez:al:96} and \citet{Frasca:00}.  Analysis of
high resolution \euve\ data
\citep{Griffiths:98,Brickhouse:Dupree:al:99,SanzForcada:al:03} showed
a corona dominated 
by material at $T\sim 10^{6.9}\,$K, and a substantial amount of
material at hotter temperatures of $T\sim 10^{7.3}\,$K.

\section{Observations and Data Processing}\label{sec:obs}

\subsection{\chan/\hetgs}\label{subsec:cxoobs}

We observed AR~Lac six times (observation identifiers (\obs), 6--11)
with the \chan\  X-Ray Observatory's \hetg/ACIS-S instrument (\hetgs)
\citep{Weisskopf:02} in standard timed-event mode.  Two longer
exposures (35 ks) were at the same orbital quadrature.  Four shorter
observations ($\sim8$ ks) were paired at the two
eclipses.\footnote{Caveat: due to an 0.5 day ephemeris error, comments
  on the phase in the standard data product headers are wrong.}  In an
attempt to minimize uncertainties from any long term trends in
activity, all the data were taken within five orbits of this two day
period binary.  We applied the ephemeris of \citet{Perryman:97a},
which differs from the more recent determination by \citet{Marino:98}
by less than 0.01 in phase at the epoch of observation.  Observational
details are given in Table~\ref{tbl:obsdata}.

The event files were re-processed to apply updated calibration files
(namely, CCD gain and bad-pixel filters; we used {\tt ASCDSVER CIAO
  2.1 Wednesday, February 28, 2001} and {\tt CALDBVER} 2.3).  Data were
also ``de-streaked'' to filter out the instrumental artifact on CCD-8
and then processed to grating coordinates.  These event lists were
then binned onto the standard \ciao\footnote{\tt http:cxc.harvard.edu}
\hetgs\ spectral grids.  Effective area tables (auxiliary response
files, or ARFs) were made with \ciao\ software ({\tt mkgarf}) for each
of the \heg\ and \meg\ gratings, and for $+1$ and $-1$ orders.  We
show a summary flux spectrum in Figure~\ref{fig:flux}.

\subsection{\euve}\label{subsec:euveobs}

\euve\ spectrographs cover
the spectral range 70--180~\AA, 170--370~\AA\ and 300--750~\AA\ for
the short-wavelength (SW), medium-wavelength (MW) and long-wavelength
(LW) spectrometers respectively, with corresponding spectral
dispersion of $\Delta\lambda\sim 0.067$, $0.135$, and $0.270$
\AA/pixel, and an effective spectral resolution of
$\lambda/\Delta\lambda$$\sim200$--$400$. The Deep (DS) Survey Imager
has a band pass of 80--180~\AA\ \citep{Haisch:Bowyer:al:93}.
Standard data products from the \euve\ observations of AR~Lac were
obtained through the 
Multimission Archive at Space Telescope (MAST), corresponding to three
observational campaigns starting in 1993 October 12 (96~ks), 1997 July
3 (74~ks) and 2000 September 14 (63~ks).

\section{Photometric Analysis}\label{sec:corstruct}
\subsection{Lightcurves}

We made light curves of the \chan\ data using 
the \ciao\ program {\tt lightcurve}, and
filtered the input events so that $-3$ to $+3$ orders (excluding
zero, which is piled\footnote{``Pileup'' refers to coincidence of
  photons in the same spatial and temporal bins. It makes the response
  non-linear, and also, if severe, censors counts via on-board
  filtering.}),  
and \meg\ and \heg\ photons within 1--25 \AA\ were all binned
into one curve for each \obs.  We also made light curves in some
stronger line and continuum bands by filtering on narrow wavelength
regions. We show some of the light curves in Figure~\ref{fig:lc}.

\euve\ light curves (small open squares in Figure~\ref{fig:lc}) were
built from the DS image by taking a circle centered on the source, and
subtracting the sky background within an annulus around the center.
Standard procedures were used in the IRAF package EUV v.~1.9, with a
time binning of 600~s.  In the following discussion and analysis, the
observations in 1993 and 1997 are referred to as ``quiescent'' states;
despite of the presence of some minor flares the apparently quiescent
component dominates the integrated flux \citep{SanzForcada:al:03}.
The observations in 2000 are 
instead referred to as a ``flare'' state.  Only the observations of
the 2000 campaign, contemporaneous with the \chan\ observations, are
employed here.

\subsection{Variability}

\begin{figure}
  \figurenum{2}
  \epsscale{0.9}
\plotone{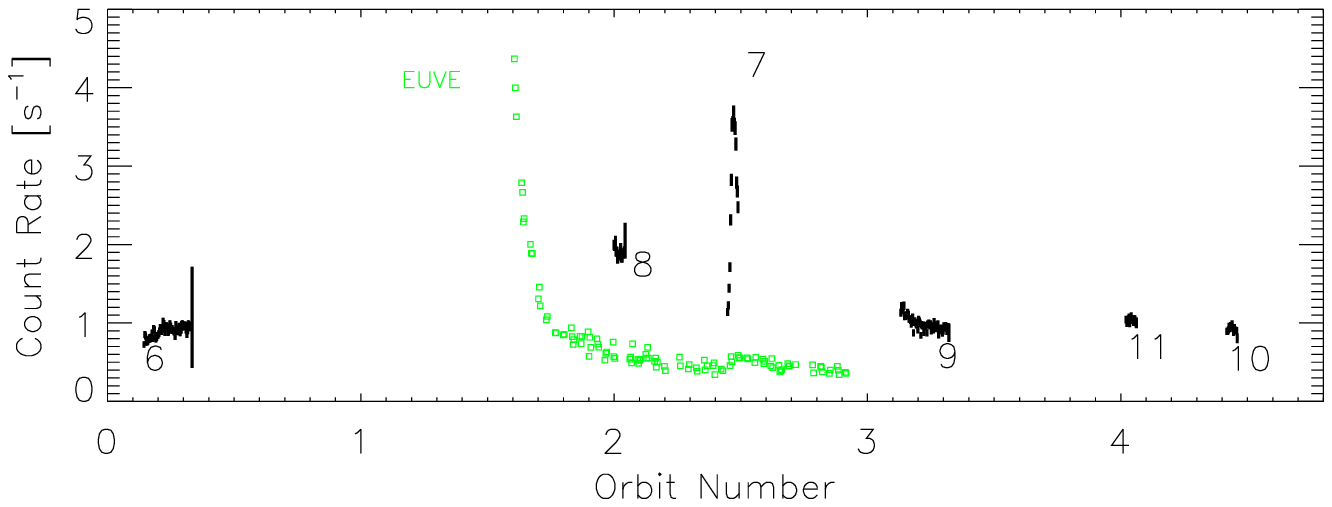} 
\plotone{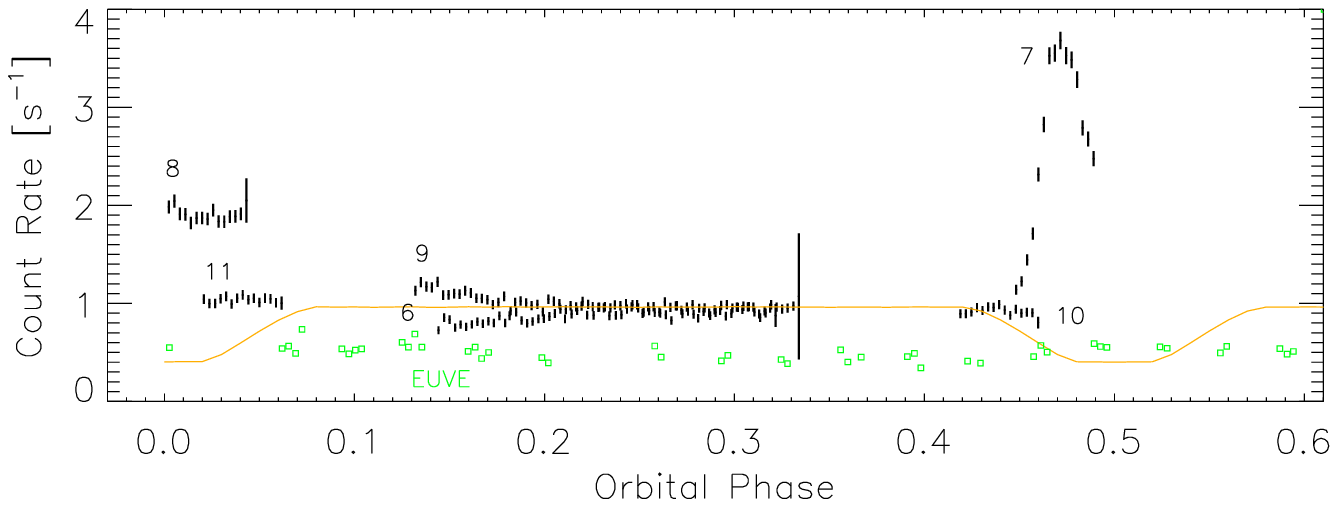} 
\plotone{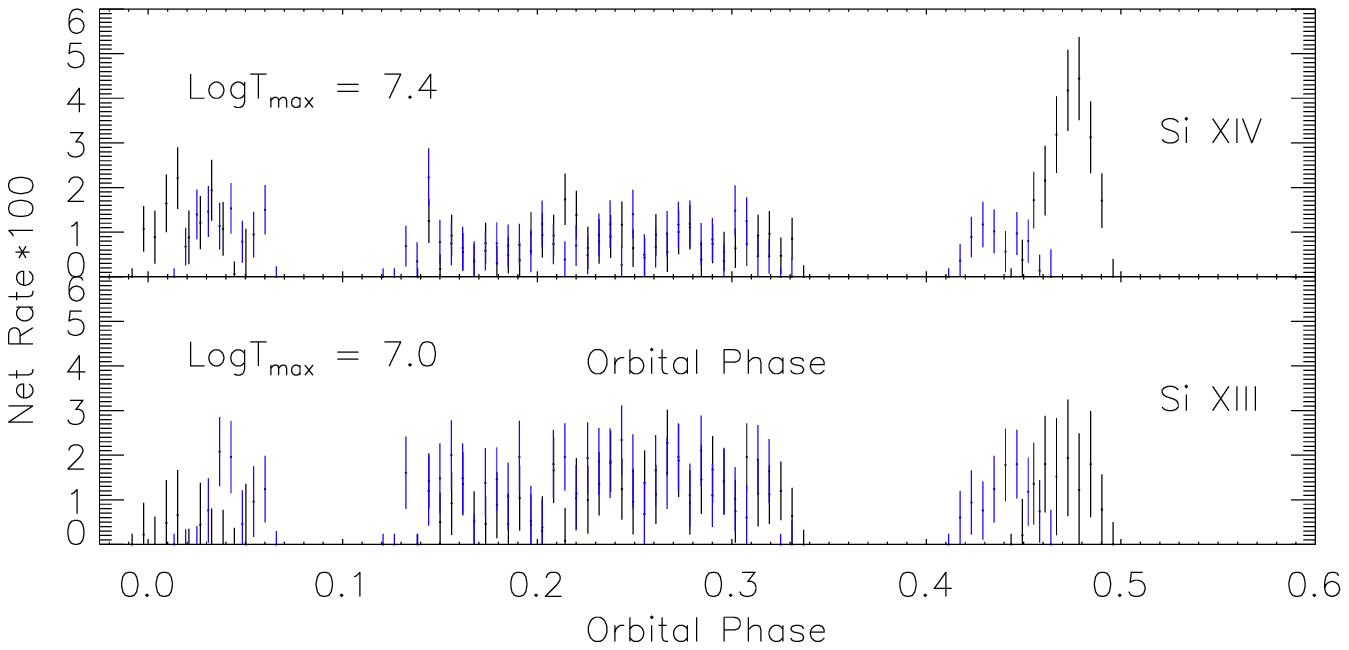} 
  \caption{}
\label{fig:lc}
\end{figure}

From the X-ray light curve in Figure~\ref{fig:lc} we see that \arl\ 
was highly variable during the observations.  There is one obvious
flare near phase 0.5 (orbit 2.5, \obs\ 7) when the count rate
increased by nearly a factor of four, and then rapidly decayed.  
A little more than half an orbit later (orbit 3, phase
0.13-0.33, \obs\ 9), the count rate was decreasing.  It is
tempting to associate this with the tail of the larger flare, but it could
just as well be an independent event.   For the two consecutive
segments to connect smoothly, the decay rate would have to be variable
in a complicated way; the estimated decay rate of \obs\ 9 is too short
to connect with the previous flare.  Given that \obs\ 8 is somewhat
elevated in count rate and fairly steady, this would be consistent with 
the high flare frequency and broad range of decay rates that 
\arl\ typically exhibits, as revealed by the extensive EUVE and Chandra
lightcurves analysed by Pease et al.\ (2002).

We can use the \euve\  light curve (Figure~\ref{fig:lc})
to provide a context for the disjoint X-ray observations.
The \chan\  \obs\ 7 flare was simultaneously observed with the \euve,
which showed a small step up in count rate.  A large EUV flare
occurred between \obs\ 6 and 8.  Assuming that the high X-ray to EUV
enhancement ratio seen in \obs\ 7 always holds, which would be 
more indicative of
further heating of an existing volume of hot plasma rather than
evaporation of cold chromospheric material, the \euv\ flare must
have started after \obs\ 6.  The baseline trend from \obs\ 8-7-9 is
probably indicative of the decay of the large flare.
There is no obvious quiescent state; the nearest hint of one is from
phase 0.25-0.30. One observation (\obs\ 9) is decreasing to this
level, but the other (\obs\ 6) is actually increasing from a lower
flux state.

There are no obvious X-ray eclipses, but exposures were short at these
phases.  The dotted curve in Figure~\ref{fig:lc} is a simple
occultation model with uniform disks of equal surface brightness and
with relative radii in proportion to the \arl\ components, scaled
arbitrarily.  The lower X-ray flux state during primary eclipse egress
(\obs\ 11; phase 0.0-0.1) is constant.  This implies that the
emerging, smaller G-star is X-ray dark, is dark on the portion being
exposed, or that emission structures are large and polar so they are
not occulted.  We will rule out the former case later, based on
line-profile information (see~\ref{sec:lineshapes}).

To investigate qualitative temperature changes, we binned light curves
in narrow bands, for both continuum and strong line regions.
Continuua near lines were used to derive net line rates, and some
features with similar temperatures of maximum emissivity were summed
to improve statistics.  An example is shown in Figure~\ref{fig:lc} for
\eli{Si}{14} (6.18 \AA, $\log T_{max}=7.4$), which follows the overall
integrated light 
curve, and \eli{Si}{13} (6.6-6.8 \AA, $\log T_{max}=7.0$), whose net
rate shows little 
variation.  As a qualitative temperature diagnostic, we computed
the modulation in the count rate, $r$, for each feature, defined as
$(r_\mathrm{max}-r_\mathrm{min}) / (r_\mathrm{max}+r_\mathrm{min})$,
which is 1.0 for maximum modulation ($r_\mathrm{min}=0)$, to 0.0 for
constant count rate ($r_\mathrm{max}=r_\mathrm{min}$).  In
Figure~\ref{fig:lcmod} we show the modulation as a function of
temperature of formation.  For lines, the temperature of formation is
defined as the temperature of maximum emissivity.

The trend in modulation is clear from these light curves: higher
temperature emission is more modulated.  The cutoff is quite sharp for
the lines at about $\log T = 7.0$.  The ion temperature, though, is
not a unique diagnostic.  For example, the difference in modulation
between \eli{Mg}{12} and \eli{Si}{13} is not contradictory, since
\eli{Mg}{12} is hydrogen-like and has a long emissivity tail extending
to higher temperatures.  The continuum modulation showed a similar
trend, in that the continuum flux at shorter wavelengths, which is
formed by higher temperature plasma, is also more strongly modulated.

As another qualitative diagnostic of temperature changes, we looked at
band light curve ratios to see if the decrease in \obs\ 9 is a flare
decay (after the \euve\  observation, on the extreme tail of the EUV
flare), and whether the slight rise in \obs\ 6 in the same phase
interval (but before the EUV flare) could be due to rotational
modulation of asymmetrically distributed X-ray emitting structures.
We used the strongly modulated 1.9-2.9 \AA\ band light curve for the
high temperature diagnostic, the relatively unmodulated \eli{Fe}{17}
regions' continuum bands (14.7-14.9, 16.4-16.6, and 17.2-17.5 \AA) as
well as \eli{Ne}{10} to sample the cooler plasma.  The ratios clearly
showed the flares near phases 0 and 0.5, but no significant
differences at the quadrature phases.  We cannot say whether either
the slow rise or fall are due to flares or rotational modulation.  But
given the large EUV flare, the \obs\ 9 decrease is probably due to
flare decay, and temperature changes are below the sensitivity of the
observations.

\section{Spectroscopic Analysis}
\subsection{Spectral Line Fluxes}
\subsubsection{Chandra Spectra}

Obtaining line fluxes from coronal X-ray spectra such as that of \arl\
is complicated by the presence of continuum flux.  This continuum has
a shape which is dependent primarily on the plasma temperature.  In
order to extract line fluxes, we therefore adopted an iterative
approach, whereby temperature information from spectral lines was used
to calculate a model continuum, which was then used to refine
measurements of the line fluxes.

We summed $\pm1^\mathrm{st}$ orders and fit line fluxes with the
\cxc\ software suite, \isis\ \citep{Houck:00}.\footnote{\tt
  http://space.mit.edu/CXC/ISIS} Emission lines were fit by convolving
intrinsic source line profiles (Gaussians) plus a model plasma continuum by
the instrumental response or line spread function (LSF). 
The free parameters were the 
Gaussian centroids and areas of each line, and, if necessary, the Gaussian
dispersion.  Typically, the lines are unresolved, so the Gaussian
dispersion was frozen at a value well below the instrumental
resolution.  For some cases, either due to blends or to orbital
velocity separation, we fit the line width (line widths are discussed
in \ref{sec:lineshapes}).  The redistribution component of the
response for grating spectra is the line spread function (LSF), which
is stored in grating redistribution matrix files (RMF) in the \ciao\ 
calibration database.  \heg\ and \meg\ spectra were kept separate, but
fit simultaneously.  The continuum model was obtained iteratively,
first by fitting relatively line-free regions with a single
temperature plasma model, and subsequently by using the result of
differential emission measure models to improve the predicted
continuum.  The continuum normalization was not allowed to vary in the
fitting since the apparent continuum is often significantly above the
true continuum due to line blending.  For the continuum model, we use
the summed true continuum and pseudo-continuum components in the
Astrophysical Plasma Emission Database \citep[\aped;][]{Smith:01}.

To be explicit, if the model line contribution, $S$, to a region is
expressed as a sum of normalized Gaussians, $g$, enumerated by
component $i$, as
\begin{equation}
  \label{eq:linegauss}
  S(\lambda) = \sum_i{a_i\,g(\lambda-\lambda_i, \sigma_i) },
\end{equation}
then the $n^\mathrm{th}$ iteration predicted counts spectrum in the
region of interest can be defined as
\begin{equation}
  \label{eq:linecounts}
  C(h) = \int_{\Delta\lambda}
          { d\lambda\, R(\lambda,h)\, A(\lambda) 
    \left[ S(\lambda) + S_{c}^{n}(\lambda) \right]}. 
\end{equation}
Here, $h$ is the detected channel, $\lambda$ the wavelength, $S_c$ the
continuum source model, $R$ the redistribution function (LSF; or
response matrix, RMF), and $A$ the effective area. The $a_i$ values
are the line fluxes, and $\lambda_i$ and $\sigma_i$ are the line
centroid and dispersion, respectively.  The line fluxes determined by
minimizing this against the counts are listed in
Table~\ref{tbl:lineflux}.

\subsubsection{EUVE Spectra}

Spectra of AR~Lac, binned over the total observation, were used to
provide fluxes of lines in the range $\sim$90--400~\AA\ as listed in
Table~\ref{tbl:lineflux}.  To correct the observed fluxes for
interstellar hydrogen and helium continuum absorption, we used a ratio
\eli{He}{1}/\eli{H}{1}=0.09 \citep{Kimble:Davidsen:al:93}, and a value
of $N_H=1.8\times10^{18}$, calculated from the \eli{Fe}{16} $\lambda
335/361$ line ratio.  Lines of \eli{Fe}{9}--\eli{Fe}{24} (except
\eli{Fe}{17}) are formed in this spectral range, providing good
coverage in temperature from only one element over the range
$T\sim10^{5.8}-10^{7.4}\,$K, avoiding the introduction of further
uncertainties from the calculation of abundances.

\subsection{Temperature Distribution}

A more quantitative description of plasma temperatures is given by the
emission measure distribution.  This is a  one-dimensional
characterization of an emitting plasma describing the emitting power
as a function of temperature.  It does not tell us how the material is
arranged geometrically.  This must be derived from other information,
such as eclipse or rotational modulation, or inferred through
(hopefully realistic) assumptions of parametric or semi-empirical
models such as hydrostatic, magnetically confined loops.  Nonetheless,
the emission measure distribution remains a useful quantity for
visualization and comparative study of coronal temperature structure.
\citet{BowyerDrake:00} present a good review of emission measure
modeling in the context of cool stars and extreme ultraviolet spectroscopy.

There are many pitfalls in inverting the emission integral
\citep{Craig:76, Hubeny:95, McIntosh:98, Judge:97}.  Inversion is
limited by the emissivity functions of real atoms, which do not form an
orthogonal set of basis functions; some kind of regularization is
necessary.  \citet{Kashyap:98} also discuss problems of spurious
structure caused by errors in atomic data, and describe an approach to
the problem based on a monte carlo technique converged by a Markov
chain which includes atomic data as well as line measurement uncertainties.

In this analysis, we fit the differential emission measure (\dem) and
abundances simultaneously by minimizing the integrated line flux
residuals using the emissivities from the \aped\ and the ionization
balance of \citet{Mazzotta:98}.  Our basic method is described by
\citet{Huenemoerder:01}.  We minimize a statistic, %
\begin{equation}
   \label{eq:demchisq}
   \chi^2 = \sum_{l=1}^{L}
             \frac{1}{\sigma_l^2}
             \left[f_l - A_{Z(l)}\frac{\Delta\log T}{4\pi d^2}
             \sum_{t=1}^{N}{\epsilon_{lt}\Psi(e^{\ln D_t},k)} \right]^2
\end{equation}
Here, $l$ is a spectral feature index, and $t$ is the temperature
index. The measured quantities are the line fluxes, $f_l$, with
uncertainties $\sigma_l$. The {\em a priori} given information are the
emissivities, $\epsilon_{lt}$, and the source distance, $d$.  The
minimization provides a solution for the differential emission
measure, $D_t$ and abundances of elements $Z$, $A_Z$.  The
exponentiation of $\ln D$ is simply a trick which forces $D_t$ to be
non-negative, and $\Psi$ is a smoothing operator which imposes some
implicit regularization on the solution; we use a Gaussian convolution
with a dispersion of 0.15 dex.  We omit spectral features which are
line blends of different elements and of comparable strengths.  We use
a temperature grid of 60 points spaced by 0.05 in $\log T$, from $\log
T=5.5$ to 8.5.  The emissivities, $\epsilon$, are as defined by
\citet{Raymond:96}.
We further improve upon the method by implementing a Monte Carlo loop in
which we performed the fit 100 times, each time perturbing the
observed line fluxes by their measurement uncertainties, assuming a
Gaussian distribution.  In this way we are able to obtain a variance
on the \dem\ at each temperature as well as on each elemental
abundance according to the quality of the line measurements
(systematic uncertainties in the calibration or the atomic database
are, however, still present).  After a \dem\ was obtained, we re-fit
the lines using the plasma continuum predicted by that \dem.  This
iteration was done three times.  We also used the synthetic spectrum
to adjust the abundance and \dem\ scale factor to match the observed
line to continuum ratio, and hence the absolute abundance scale
relative to the Solar values of \citet{Anders:89}.  During the
iterations, we compared the model and observed spectra in detail to
examine various line series and were able to identify and include
weaker features, and reject some features as probable blends or
mis-identifications.
 
We show the observed and synthetic counts spectra in
Figures~\ref{fig:countsA}-\ref{fig:countsD}, the \dem\ in
Figure~\ref{fig:dem}, and list the abundances in
Table~\ref{tbl:abund}.  The line measurements and predicted fluxes are
listed in Table~\ref{tbl:lineflux}.

Our method imposes some implicit regularization by smoothing the \dem\ 
over a few temperature grid points.  In fitting all elements
simultaneously, we can couple the \dem\ over broader temperature
ranges and simultaneously derive self consistent relative abundances
and \dem.  This is necessary to fit the low temperature regions which
are not covered by the temperature ranges of the emissivities of the
iron ions 
available with \hetgs.  By including fluxes from \euv\ lines (e.g.,
\eli{Fe}{9}--\eli{}{16}), we are able to constrain the low temperature portion
of the \dem\ ($\log T < 6.4$), which \eli{N}{7} and \eli{O}{7} sample
in the \hetgs\ spectrum.  We tested consistency by fitting elements
independently or in small groups.  While fits were of lower quality
and limited in temperature range, they were consistent, so we
performed the final fit over all elements, ignoring the
density-sensitive He-like forbidden and intersystem lines, some
strong blends (e.g., \eli{Ne}{9}$+$\eli{Fe}{19} 13.4 \AA), and some
obviously discrepant features which are likely to be either
mis-identifications, compromised by blends, or lines for which
emissivities are not accurately known.

Since there are poorly quantified systematic uncertainties in the
atomic data (and thus emissivities), we repeated the \dem\ fits with a
lower limit to the flux uncertainty of 25\% (regardless of the counts)
to globally approximate the atomic data systematic errors.  The
resulting DEM structure was the same, but with appropriately larger
uncertainties.  This lends confidence that the \dem\ structure is real
and not due to deficiencies in atomic data.

At no point do we formally minimize a binned spectrum against a binned
synthetic spectrum.  This had been the only option and the norm for
low-resolution X-ray spectral modeling with previous missions such as
\asca\ and \rosat.  There are many ways for such ``global'' fitting to
fail for high-resolution spectra: inaccurate model wavelengths can lead
to a mis-match between predicted and observed lines; 
any spectral region may have features missing from the
emissivity database; emissivities may be inaccurate; or continuum bins
can dominate $\chi^2$, to list a few pitfalls.  For these reasons, we
prefer a strictly line-based analysis because we can more finely
manipulate the features to be fit according to their ionic sequences,
density sensitivity, blending, temperature of formation, or any other
parameter available in the atomic database.  These techniques have
long been in use for UV and EUV emission line spectroscopy of coronal
plasmas.

The \dem\ we obtain is dominated by two large peaks, at $\log
T\sim6.9$ and $\sim7.4$.  There is a hot tail imposed by the presence
of \eli{Fe}{25}, \eli{Fe}{26}, \eli{Ca}{20}, and \eli{Ar}{18} (within
large uncertainty, since the lines are of low signal to noise ratio)
and the short wavelength continuum.  At $\log T=6.2$, there is a weak
peak required by the \euv\ lines from lower ionization states of iron
({\sc ix-xvi}) which provide a much better constraint than does the
relatively low signal-to-noise detection of \eli{N}{7} (24.8 \AA).
The overlapping temperature distributions of N and the EUV Fe lines
serve to better constrain the abundance of N.  We note that the
derived \dem\ is similar to that obtained for the RS~CVn system
HR~1099 by \citet{Drake:01} based on \chan\ \hetg\ spectra, while the
double-peaked structure at high temperatures is also reminiscent of
the emission measure distributions for AR~Lac, HR~1099, II~Peg, and
other stars derived from EUVE spectra by \citet{Griffiths:98} and
\citet{Sanz-Forcada;Brickhouse:02}; the cool bump is also seen in some
of the emission measures derived by the latter authors.  It is
tempting to conclude that the detailed structure in the \dem\ derived
in this study and others does indeed reflect the true source
temperature structure.  We caution, however, that such apparent
structure can also arise as a result of errors in the underlying
atomic data \citep[see, e.g.,][]{Kashyap:98}.  In particular, errors
in ionization equilibria could plausibly induce spurious structure
since such errors would be highly correlated with temperature.  We
therefore emphasize that detailed interpretation of the \dem\ 
structure should proceed with caution.

The very hot portion of the \dem\ appears to be predominantly 
due to the flares,
since the modulations from \eli{Fe}{25} and the nearby continuum are
nearly 90\%.  Much of the large peak ($\log T=7.2$--$7.6$) is
contributed by flares, having a modulation greater than 50\%.  Below
7.0 there is little modulation.  This greater variability in the
hottest plasma component has been seen in other active stellar coronae
\citep{BowyerDrake:00}, such as 
II~Peg \citep{Huenemoerder:01} during flaring, in Capella
\citep{Brickhouse:00} in widely separated observations, and in the
RS~CVn stars studied with \euve\  by \citet{Sanz-Forcada;Brickhouse:02}.

\subsection{Abundances}

Relative abundances are determined by the \dem\ fitting procedure, and
are strongly correlated with the \dem\ solution.  An element which is
isolated in temperature from other ions will be degenerate in the
\dem\ normalization and abundance, since the flux is determined by
their product.  By fitting all ions simultaneously we remove some of
the degeneracy since a series of intermediary strongly overlapping
emissivities can couple ions which only overlap slightly.  By
performing a Monte Carlo iteration we derive an estimate of the range
of solutions allowed by measurement uncertainties.

The abundances we derived are listed in Table~\ref{tbl:abund}, and we
plot them against the first ionization potential (FIP) in
Figure~\ref{fig:abund}.  The abundances range from about half to about
double the Solar photospheric values, with no simple trend with FIP.
Neon is about 1.7 times the accepted Solar value and three times the
iron abundance.  This neon to iron ratio is a fairly reliable
determination since the lines form in overlapping temperature ranges
and so are relatively independent of the \dem\ (assuming they form
from ions within the same volume). If we average Al and Ca at the
lowest FIP, we find them near Solar, and significantly different from
the average of Fe, Mg, and Si.  Al and Ca form in the same temperature
range as Si, so this difference is also relatively independent of the
details of the \dem.

The photospheric abundances of AR~Lac \citep{Gehren:99} of iron, silicon,
and magnesium (all low FIP elements) are systematically
higher than the coronal values.  Taking the error-weighted means, we
find the relative abundance to be $0.6\pm0.06$ in the coronae, as
compared to $1.0\pm0.13$ in the photosphere.  The average of the
lowest FIP elements, Al and Ca, is $1.2\pm0.2$, which is comparable to
the photospheric value.

\citet{Kaastra:96} and \citet{Singh:96} analyzed the same \rosat\ and
\asca\ data with different methods, but obtained statistically
identical results for everything but neon.  Their values were about
half of ours: $0.3\pm0.02$ (again averaging over Fe, Si, Ca, and Mg).
The latter authors gave a good synopsis of the systematic
uncertainties in low resolution spectral fits and do show how changes
of 50\% can arise from fitting different spectral regions.  Both
studies also obtained values systematically lower than ours for all
other elements fit (S, O, N, Ar, Ne).
\rev{       
Global fits to low-resolution Beppo-SAX spectra of \arl\ by
\citet{Rodono:99} obtained an average metal abundance of 0.66, similar
to our values for Fe, Si, S, Mg, and O.
}

In comparison to previous measurements from low resolution data, we
are most discrepant with \citet{Singh:96} in the sulfur abundance,
which they claimed was robust even for low resolution data, and also
differ greatly in Al and Ca.  We provide a more robust nitrogen
abundance, for which they suspected large calibration systematics. 
The high FIP element abundances are near (N, Ar) to above (Ne) Solar
photospheric values; while there are no photospheric measurements for
these elements for AR~Lac, it seems likely that they follow the
approximately solar values for the elements studied by
\citet{Gehren:99}.

\subsection{Line Shapes}\label{sec:lineshapes}

Both stellar components of \arl\ have been seen to be X-ray active
\citep{Walter:83, Siarkowski:96}.  We have not been able to detect
eclipses with our limited phase coverage, but the spectral resolution
and phase coverage do permit us to search for line profile variations.
At quadrature, the radial velocity separation of the binary components
is $230\,\kms$ \citep[see][for a collation of system
parameters]{Gehren:99}.  The instrumental full-width, half-maximum
(\fwhm) of 0.02 \AA\ for \meg\ and 0.01 \AA\ for HEG yields
$316\,\kms$ at \eli{O}{8} (19 \AA) and $200\,\kms$ for \heg\ at
\eli{Fe}{17} (15 \AA).  We fit the summed quadrature spectra (\obs\ 6
and 9) with one or two instrumental profiles (no thermal or turbulent
components included), and with a single broadened Gaussian convolved
with the instrumental response.  The lines were not well fit by single
instrumental profiles.  Equally good fits could be obtained with
either a single Gaussian with a dispersion of about 0.005--0.01 \AA,
or by two Gaussians.  The \eli{O}{8} two-Gaussian fit is close to the
separation expected if both stars are active and the activity is
localized near their respective photospheres: the 90\% velocity limits
are at about the instrumental resolution and overlap the orbital
velocities, $ -162 (-196:-103) \kms$ for the G star, and $123 (55:197)
\,\kms$ for the K star (numbers in parenthesis are 90\% confidence
limits).  These values are in reasonable agreement with the orbital
velocities of $-115$ and $115\,\kms$, respectively, with a $30\,\kms$
range within the observation.

To test for broadening in a way which does not depend on the
calibration of the instrumental profile, we compared line profiles
between the quadrature and conjunction phases.  Both the \meg\ 
\eli{O}{8} and \heg\ Ne-Fe 12\AA\ blend were broader than the
instrumental profile at quadrature, and were consistent with the
instrumental profile at eclipses.  We show the profiles and
differences in Figure~\ref{fig:broadlines}.

\subsection{Density}

The helium-like triplet lines are well known density diagnostics
\citep{Gabriel:69,Gabriel:73,Pradhan:81,Porquet:00}.  The critical
densities increase with atomic number and are sensitive over the
ranges in $\log N_e $ of about 10--12 for \eli{O}{7} ($\lambda\lambda
21.6$, 21.8, 22.1, $\log T_{max}=6.3$), 11--13 for \eli{Ne}{9}
($\lambda\lambda 13.45$, 13.55, 13.70, $\log T_{max}=6.6$), and 12--14
for \eli{Mg}{11} ($\lambda\lambda 9.17$, 9.23, 9.31, $\log
T_{max}=6.8$), which span ranges of interest for coronal plasmas
(wavelengths refer to the resonance, intersystem, and forbidden lines,
respectively).  The density determination depends upon flux ratios
including the weak intersystem lines.  Positive detection requires high
signal, accurate continuum, and resolution of blends.  
If we examine the spectra (Figure~\ref{fig:counts}), we see that the
continuum is fairly well modeled, so this is not the limiting factor
in our measurements.  The model significantly underestimates the
\eli{Ne}{9} (Figure~\ref{fig:countsB}) and marginally overestimates
\eli{O}{7} (Figure~\ref{fig:countsD}) forbidden lines. The
\eli{Mg}{11} resonance and forbidden lines match well
(Figure~\ref{fig:countsB}).

None of our ratios give a good density constraint: \eli{O}{7} is weak,
\eli{Ne}{9} is seriously blended with Fe and is listed here only for
completeness, and the \eli{Mg}{11} intersystem line is weak and
possibly blended with high-$n$ \eli{Ne}{10} H-like transitions.  The
formal ratios indicate logarithmic densities ($\mathrm{cm\mthree}$)
of about 10.8 from oxygen, but with 90\% uncertainties which span the
range from 9--12; for neon, 11 with 90\% uncertainties up to 11.5; for
magnesium, 12.2 with a 90\% upper limit of about 12.8.  Neon and
magnesium can be considered to define upper limits, but the lower
limits are unconstrained.  The neon intersystem line could be up to
about 30\% iron blends \citep[][Ness et al., 2003, in preparation]{Ness:02}, which would
lower the upper limit 
somewhat.

Hence, we tentatively
conclude that we have detected densities on the order of $\log N_e \sim
11$, high enough to affect \eli{O}{7} and \eli{Ne}{9} ratios, but not
\eli{Mg}{11}.  There is no reason, however, for the values to be
identical for different ions, since they form at different
temperatures.  Without better measurements of the intersystem lines,
we do not have data to constrain $N_e(T)$ more rigorously.

\section{Discussion}

The high resolution X-ray spectrum of \arl\ provides a wealth of new
information about the system.  
While we have derived more detailed emission measure
models and abundances than possible from low resolution data, there
are still inadequacies in the models and significant problems to be
solved.  We must remember that there are two stars' coronae involved, and
that they are highly variable, making comparison with
other epochs difficult.  The composite nature is not a problem for
derivation of emission measures since that quantity does not assume
any geometric structure.  Interpretation of the emission measure,
however, does require information or assumptions about the geometrical
distribution of plasma.

The line widths, being barely resolved by \hetgs, show that both stars
are active, as has long been known from the first X-ray observations
\citep{Walter:83} as well as from spectroscopy in the ultraviolet
\citep{Neff:89,Pagano:01} and optical \citep{Frasca:00}.
Spatial structure can be derived from line velocities or occultations.
The only conclusive modulation is from flares, which have
high enough frequency and duration \citep[also noted by][for
\eli{Mg}{2}]{Pagano:01} to make it very difficult to infer spatial
structure.  Prior studies suggested the G-star to have a compact and
uniform chromosphere, and compact but highly structured corona
\citep[eg]{Pagano:01}.  The current lack of modulation during primary
eclipse egress coupled with line broadening at quadrature is
consistent with this view.  The modulation seen by
\citet{Siarkowski:96} seems to be ruled out; their \asca\ light curves
showed a deep and long primary minimum, which they modeled as extended
emitting structures between the two stars.  Whether such extended
material is indeed present will require better phase coverage at
different epochs as well as comparison to coronal models in stars
which should not have binary interaction.  The high frequency and
amplitude of flare variability, however, may mean that image
reconstruction of the corona is not feasible \citep[Drake et al.,
2003 (in preparation)]{Pease:Drake:al:02}.

The \dem\ and abundance model does predict the spectrum reasonably
well (see Figure~\ref{fig:counts}), but there are still some very
large discrepancies for individual lines.  Some iron line intensities
are well modeled, but others exhibit significant discrepancies.  For
example: \eli{Fe}{18} $\lambda 14.208$ is significantly stronger in
the model, while the neighboring blend of \eli{Fe}{18}
$\lambda14.256$, \eli{Fe}{20} $\lambda 14.267$ is well modeled in the
predicted two-to-one ratio.  The \eli{Fe}{17} lines, as a series, are
also poorly fit: $\lambda 17$ pair model is slightly weak, $\lambda
16.78$ model is slightly strong, while the $\lambda 15.01$ model is
nearly twice as strong as the data.  This latter discrepancy was also
noted for II~Peg \citep{Huenemoerder:01}.  Instead of interpreting the
apparent weakness of the $\lambda 15.01$ line as opacity, we suspect
that the emissivities may need revision.  The new Fe~L-shell
calculations of \citet{Gu:02} and by \citet{Doron:Behar:02} indicate
that some ratios may differ from earlier calculations by about a
factor of 1.5.  We are working to include updated emissivities into
the synthetic spectrum.

There could also be deficiencies in the \dem\ and abundance model.  We
could increase the model strength of \eli{Fe}{17} $\lambda 15.01$ by
increasing the \dem\ on the low-temperature side of its emissivity
distribution ($\log T\sim 6.5$--$6.6$), so as to minimize the affect
on \eli{Fe}{18}.  This, however, would have a ripple of side affects:
the abundance of Mg and Ne would have to be reduced, but enhancing the
cool \dem\ would also change the ratios of H-like \eli{Mg}{12} and
\eli{Ne}{10} to their He-like states, since the H-like emissivities
span the hot peak of the \dem.  The emissivities of neon, in turn, overlap
significantly with those of oxygen, and so forth.  We have already implicitly
optimized the balance of abundances in the minimization.  However,
there may be local minima, or our solution may be skewed by erroneous
iron emissivities, or may be too smooth to find very sharp temperature
structure.  Line blending is also a significant problem, and will
require a non-local approach in which isolated lines of a blending
species are used to
predict the contribution of that species to blended features. We have
also assumed that each stellar binary component has the same
composition, which might be reasonable but is unverified. Future efforts will
address these issues.

The normalized line flux residuals ($\chi = (data-model)/\sigma$) are
plotted in Figure~\ref{fig:chiflux}.  The scatter is quite large, with
$\chi^2/\nu\sim3$.  The \eli{Fe}{17} lines form near $\log
T=6.7$--$6.8$, and we can see that we have lines both stronger and
weaker than the model, as well as a number which are well fit.  In
addition to the significance of deviations, the bottom panel of the
figure shows the flux ratio to the model, which indicates the
percentage deviations.  We note that the reduced $\chi^2$ of the
spectral energy distribution itself is about 1.0; this is misleading,
because it is dominated by a large number of well modeled continuum bins
with relatively low signal-to-noise ratios.  This illustrates the
problem that minimization of a model
against a binned spectrum instead of against just the extracted line
fluxes places less statistical weight on spectral lines thought to be
useful diagnostics, and so devalues the line flux information.

There is at least one other potential source of error in the \dem\
modeling.  We have assumed collisional ionization equilibrium (CIE),
in which collisional excitation and ionization from the ground state,
followed by radiative decay, recombination, and di-electronic
recombination are the dominant processes.  We have previously argued
\citep{Huenemoerder:01} that this is acceptable even if there is a
large flare contribution to the flux, since most lines are expected to
thermalize quickly \citep{Golub:89,Mewe:85,Doschek:80}.  That CIE is
valid in {\em all} regions of the plasma, though, is an assumption.
If flaring is very frequent, or if the apparently quiescent emission
were to arise as a result of a superposition of many unresolved,
weaker flares, then the plasma might be driven out of CIE.  In this
regard, improving the model is not a trivial proposition, requiring 
a detailed attention to
each ion series in conjunction with quality assessment of model
emissivities and more rigorous incorporation of atomic data
uncertainties (including uncertainties in the ion populations) to
determine whether or not the data are indeed well-described by the
standard optically-thin plasma in thermal equilibrium.  
We will accept the current \dem\ and model parameters as the best
available, given these caveats.

The abundance trends found in other systems show that low FIP species
are sub-Solar, while high FIP elements (neon) which have enhanced
abundance
\citep{Huenemoerder:01,Drake:01,Audard:01a,Brinkman:01,Gudel:01a,Gudel:01b}.
However, we find that the lowest FIP elements have intermediate
abundances, as does the high FIP element Ar, relative to the other
elements. \arl\ seems to be more moderate in terms of abundance
anomalies compared to other active stars like II~Peg or HR~1099, being
less deficient in iron, and having a lower neon abundance ratio to
iron \citep[see][for collections of \hetgs\ spectra qualitatively
ordered by iron to neon abundance
ratio]{Kastner:02,Huenemoerder::2002}.

The trend of increasing abundance with FIP is opposite what has been
seen in the Solar corona \citep[see, e.g.,][for
reviews]{Feldman:Laming:00, Feldman:92, Feldman:90}.  FIP-based
fractionation is believed to occur in the stellar chromosphere where
low FIP species are predominantly ionized while high FIP species
remain neutral.  However, there is as yet no widely accepted
quantitative explanation of the solar FIP effect.  The new generation
of stellar observations from \chan\ and XMM-Newton have provided new
challenges for future models.  The abundances derived here for \arl\
further complicate the picture with enhanced abundances for elements
with both low and high FIP, relative to intermediate FIP elements.

The integrated emission measure we obtain for the range, $\log
T=6.0$--$8.5$, of $1.2\times10^{54}\,\mathrm{cm^{-3}}$, is similar to
previous determinations.  About one third of the emission measure is
in the peak at $\log T=6.9$, about half in the peak at 7.4, 15\% in
the hot tail, and about 2\% in the cool bump at 6.2 (note that
Figure~\ref{fig:dem} plots the \dem\ integrated over intervals of
$\Delta\log T = 0.05$).  \citet{Griffiths:98} derived emission
measures from UV, EUV, and X-ray data, and \citet{Rodono:99}, who
analyzed SAX data, found temperature components similar to our peak
\dem\ temperatures, and integrated emission measures of about half our
value. \citet{Singh:96} also derived similar temperature components
and emission measures from \rosat\ and \asca\ spectra, with an
integrated emission value more similar to ours.  These diverse
observations and analyses over several epochs are remarkably similar,
and indicate that the mean activity level of the \arl\ coronae is
relatively stable.

If we interpret the emission with a very simple geometric model of a
semi-toroidal loop of constant cross section, we can derive a loop
height as
%
\begin{equation}
      h =
      \pi^{-4/3}\,N_{100}^{-1/3}
      \alpha_{0.1}^{-2/3}
      (VEM)^{1/3} N_{e}^{-2/3} R_{*}^{-1},
\end{equation}
%
%
in which $N_{100}$ is the number of identical loops divided by 100,
$\alpha_{0.1}$ is the ratio of loop radius to length divided by 0.1,
$VEM$ is the volume emission measure, $N_e$ is the electron density,
and $R_*$ is the stellar radius \citep[see][for a
derivation]{Huenemoerder:01}.  If we attribute half the emission to
each star and use a density of $1\times10^{11}\mathrm{cm^{-3}}$, then
the loop heights are 0.08 and 0.04 (in stellar radii, relative to the
G- and K-star components, respectively).  In other words, the loops
are compact; even if there are only 10 loops instead of 100, they only
double in height, but if density were also ten times lower, then the
extent becomes significant.  If we attribute the hot part of the \dem\ 
to a single flare loop (since ``hot'' flux is highly modulated), then
that loop height is about 0.2--0.9 times the radius of the G-star, for
density values of $0.1$--$1\times10^{11}$; this is significant
compared to the size of the binary system (2--15\% of the semi-major
axis).  The simple assumptions of semi-toroidal loop intersecting a
planar atmosphere are not valid if loops are large relative to the
stellar radius, but this heuristic argument supports the existence of
extended coronal structures as has been suggested by several authors
\citep{Walter:83,Siarkowski:96,Frasca:00,Pagano:01,Trigilio:01}.
\rev{       
  From radio interferometry, \citet{Trigilio:01} derived a scale
  for the ``core'' component of the radio corona of about 0.4
  G-star radii, similar to our low-density, few-loop case.  Their
  ``halo'' component was somewhat more extended, to several stellar
  radii.  It is, however, difficult to make a meaningful comparison
  between the X-ray and radio extents due to the very different model
  assumptions and the very uncertain X-ray plasma density
  determination, and because the radio and X-ray emission to not
  necessarily originate from the same plasma.
}

In contrast to the spectral behavior, the light curves at various
epochs can be quite different.  The SAX light curve \citep{Rodono:99}
showed much structure, with a rotational modulation, a short eclipse,
and frequent and short flares.  In contrast, the \asca\ light curve
\citep{Siarkowski:96} was relatively smooth and showed broad and deep
primary eclipse and broad, shallow secondary eclipse.
\citet{Rodono:99} summarize and compare many of the X-ray
observations.  Some aspects of the data are quite confusing and
challenging to explain.  For example, the SAX data show a narrow dip
during primary eclipse, beginning {\em after} second contact (G-star
photosphere fully eclipsed), and returning to the higher level {\em
  before} third contact (G-star photosphere fully egressed).  This is
not consistent with their interpretation that the G-star corona is compact,
since the feature would have to be trailing the G-star to make ingress
constant in flux, yet leading on egress.  Instead, there must be other
emerging or erupting structures which could be on {\em either} star.
Our highly non-repeatable light curves show the extreme difficulty of
interpreting variability as rotational modulation or eclipses of
stable structures, as was also noted by \citet{Pease:Drake:al:02} for EUV
variability.  

The flux of \arl\ integrated from $1.7$--$30$ \AA\ is
$3.7\times10^{-11}\,\mathrm{ergs\,cm\mtwo\,s\mone}$.  For a distance
of 42 pc, the band luminosity is
$7.9\times10^{30}\mathrm{ergs\,s\mone}$.  This also represents the
time-averaged value over the observation, which is strongly biased to
the longer quadrature integrations at lower count rate.  For short times,
the luminosity can be several times higher; the flare of \obs\ 7
emitted about $10^{35}\,\mathrm{ergs}$.
\rev{       
  This is a typical size for RS~CVn binaries
  \citep[e.g.][]{MaggioPallavicini:2000,Huenemoerder:01} but there
  have been flares two orders of magnitude larger
  \citep{Ottmann:Schmitt:94,KuersterSchmitt:1996}. Perhaps the flare
  we missed in X-rays was of such a magnitude, given the large X-ray
  to EUV enhancement ratio we see near orbit 2.4 (see
  Figure~\ref{fig:lc}, top panel).
}

\rev{    
  Detailed independent modeling of flare and non-flare states, while
  scientifically interesting in terms of temperature and abundance
  structure, unfortunately requires better statistics than we have in
  these data.  The larger X-ray flare of \obs\ 7, while significant in
  the light curve, is relatively short for \dem\ analysis --- much of
  the count-rate change is from the continuum.  The EUVE data, even
  binned over the entire observation, serve primarily to constrain the
  low temperature emission since the signal is so much weaker than the
  \hetgs\ data where lines overlap in formation temperature.  We also
  looked for spectral differences near phase 0.15 between \obs\ 6 and
  9, but signal was not adequate, even for summed line fluxes. The
  line modulation shown in Figure~\ref{fig:lcmod} is our low-signal
  proxy for the detailed model.  Time-dependent \dem\ and abundance
  modeling of flares will have to await larger events.
}

\rev{       
We might expect large flares to result in line shifts or excess
broadening due to velocity fields.  In Figure~\ref{fig:broadlines} we
show quadrature and conjunction profiles for two lines, \eli{O}{8} and
\eli{Ne}{10}, which are near the highest resolutions obtained with
\hetgs. These features, however, form at relatively low temperatures
and are relatively un-modulated by the flares (see
Figure~\ref{fig:lcmod}). We have attributed the  broadening in
\eli{O}{8} to orbital effects.  \eli{Ne}{10} is interesting because
the conjunction profile shows a marginal blueshift, and these phases
were most affected by flares.  \citet{Ayres:01} tentatively detected a
transient blueshift in \eli{Ne}{10} in HR~1099, coincident with a UV
flare; there was no corresponding X-ray flare detected.  We likewise
find the effect inconclusive due to conflicting information and
marginal data quality.  However, the tentative result is intriguing
and should foster further observations and more sophisticated
analyses. 
}

\section{Conclusions}\label{sec:conc}

We have presented an analysis of high resolution X-ray and EUV spectra
and photometry of the bright eclipsing RS~CVn system AR~Lac.  While
our results are qualitatively similar to those of earlier studies, the
high resolution X-ray line spectra have allowed us to obtain a
significantly more detailed glimpse of the coronal temperature
structure and abundances.

The X-ray and \euv\ spectra show that the coronae of this system are
characterized by complicated structure and variability.  The overall
activity level is similar to that determined from previous
observations.  The hot portion of the \dem\ is strongly modulated by
flares, whereas the cooler portion appears relatively quiescent.  The
flare modulation is either frequent and large enough to hide
eclipses, or the dominant coronal structures are of polar origin and
are not rotationally modulated.  The steadily emitting structures
could be compact if the density estimate of
$1\times10^{11}\,\mathrm{cm^{-3}}$ is accurate, whereas if the
emission from the X-ray flare is from a single loop at this density,
then it would be significantly extended.  This ambiguity in
interpretation illustrates the importance of reliable plasma density
estimates and underscores the difficulty in obtaining such estimates,
even from high-quality \chan\ \hetg\ observations.  Given the high
resolution of the \hetgs, we have determined that both stellar
components of the system contribute significantly to the X-ray
emission. The quadrature line profiles are consistent with the K- and
G-stars being equally bright.

Coronal abundances show some similarities with those found for other
RS~CVn stars but differ substantially in detail: the highest and
lowest FIP species both appear to have higher abundances relative to
intermediate FIP ions.  For \arl, the intermediate FIP abundances are
below measured photospheric values.


\acknowledgments

We thank Pat Slane for help tracking down a half a modified-day error,
John Houck for ISIS modeling advice, Kevin Tibbetts for crunching some
data, and Dan Dewey for a critical review of a draft.  We acknowledge
support from NASA contracts NAS8-38249 and NAS8-01129 (\hetg) and SAO
SV1-61010 (CXC) to MIT (DPH, CRC); NASA contract NAS8-39073 (CXC) to
SAO (JJD); and Marie Curie Fellowships Contract HPMD-CT-2000-00013
(JS-F).


%

\clearpage

%
\begin{deluxetable}{cccc}
\tablecaption{Observational Information. \label{tbl:obsdata}}
 \tablewidth{0pt}
\tablehead{
  \colhead{\obs} &
  \colhead{Start Date \& Time} &
  \colhead{Exposure\tablenotemark{a}} &
  \colhead{Phase\tablenotemark{b}}
  }
\startdata
6&
2000-09-11T22:05:13&
32.7&
0.14--0.34\\
7&
2000-09-16T11:48:03&
7.7&
0.48--0.52\\
8&
2000-09-15T14:18:04&
9.5&
0.98--0.02\\
9&
2000-09-17T20:02:12&
32.5&
0.15--0.35\\
10&
2000-09-20T09:28:11&
7.5&
0.42--0.46\\
11&
2000-09-19T14:31:14&
7.5&
0.02--0.06\\
\enddata
\tablenotetext{a}{Exposure time is in ks.}
\tablenotetext{b}{Phase is computed for the exposure start and stop
   times using the ephemeris of \citet{Perryman:97a}. At phase 0.0,
   the G-star is totally eclipsed.}
\end{deluxetable}

\clearpage 
%
%
%
\begin{deluxetable}{rlcrrrrrr}
\tabletypesize{\scriptsize}
\tablecaption{Emission Line Data.\label{tbl:lineflux}}
\tablewidth{0pt}
\tablehead{
    \colhead{Mnemonic\tablenotemark{a}}&
    \colhead{Ion}&
    \colhead{$\overline{\log T}$\tablenotemark{b}}&
    \colhead{$\lambda_t$\tablenotemark{c}}&
    \colhead{$\lambda_o$\tablenotemark{d}}&
    \colhead{$f_l$\tablenotemark{e}}&
    \colhead{$f_t$\tablenotemark{f}}&
    \colhead{$\delta f$\tablenotemark{g}}&
    \colhead{$\delta\chi$\tablenotemark{h}}
    }
\startdata
                          Fe26HLa&   Fe XXVI &  8.1&      1.781&      1.782 (156.2)&      3.48    (4.43)&     4.40&    -0.92 &      -0.2\\
                         Fe25HeLa&   Fe XXV  &  7.8&      1.861&      1.859 (  1.9)&     23.70    (7.14)&    25.95&    -2.25 &      -0.3\\
                          Ca20HLa&   Ca XX   &  7.8&      3.021&      3.020 (110.5)&      1.97    (1.19)&     1.90&     0.06 &       0.1\\
                        Ca19HeLaB&   Ca XIX  &  7.5&      3.198&      3.192 ( 18.1)&      8.10    (2.74)&     8.37&    -0.27 &      -0.1\\
                         Ar17HeLb&   Ar XVII &  7.4&      3.365&      3.365 (  4.3)&      3.63    (1.89)&     1.16&     2.47 &       1.3\\
                          Ar18HLa&   Ar XVIII&  7.7&      3.734&      3.732 (  4.9)&      4.44    (2.08)&     5.57&    -1.13 &      -0.5\\
                        Ar17HeLar&   Ar XVII &  7.4&      3.949&      3.950 (  2.4)&      9.32    (2.49)&     8.91&     0.41 &       0.2\\
                        Ar17HeLai&   Ar XVII &  7.3&      3.968&      3.969 (  5.6)&      5.49    (2.36)&     2.32&     3.17 &       1.3\\
S16HLb,Ar17HeLaf\tablenotemark{i}&   Ar XVII &  7.5&      3.992&      3.994 (  3.5)&      6.10    (2.19)&     6.12&    -0.02 &      -0.0\\
                           S16HLa&   S  XVI  &  7.6&      4.730&      4.730 (  0.9)&     23.60    (3.42)&    26.05&    -2.45 &      -0.7\\
                         S15HeLar&   S  XV   &  7.2&      5.039&      5.039 (  1.0)&     24.85    (3.56)&    26.65&    -1.80 &      -0.5\\
                         S15HeLai&   S  XV   &  7.2&      5.065&      5.065 ( 13.8)&      6.45    (2.80)&     5.83&     0.61 &       0.2\\
                         S15HeLaf&   S  XV   &  7.2&      5.102&      5.100 (  2.4)&     13.59    (3.13)&     8.78&     4.81 &       1.5\\
                          Si14HLb&   Si XIV  &  7.4&      5.217&      5.219 (  2.2)&     12.83    (3.30)&     8.81&     4.02 &       1.2\\
                          Si14HLa&   Si XIV  &  7.4&      6.183&      6.181 (  0.4)&     68.89    (2.94)&    62.34&     6.55 &       2.2\\
                        Si13HeLar&   Si XIII &  7.0&      6.648&      6.648 (  0.5)&     44.03    (2.28)&    51.71&    -7.68 &      -3.4\\
                        Si13HeLai&   Si XIII &  7.0&      6.687&      6.686 (  1.5)&      8.49    (1.53)&     9.54&    -1.05 &      -0.7\\
                        Si13HeLaf&   Si XIII &  7.0&      6.740&      6.740 (  0.3)&     30.20    (1.99)&    20.64&     9.56 &       4.8\\
                          Mg12HLb&   Mg XII  &  7.2&      7.106&      7.105 (  0.7)&     13.47    (1.47)&    14.14&    -0.67 &      -0.5\\
                        Fe24w7.17&   Fe XXIV &  7.4&      7.169&      7.170 (  0.7)&     13.84    (1.45)&     4.33&     9.51 &       6.6\\
         Al13HLa\tablenotemark{j}&   Al XIII &  7.3&      7.172&      7.170 (  0.5)&     11.03    (0.94)&    12.00&    -0.97 &      -1.0\\
                        Al12HeLar&   Al XII  &  7.0&      7.757&      7.758 (  1.9)&     7.84    ( 1.35)&     5.51&     2.33 &       1.7\\
                         Mg11HeLb&   Mg XI   &  6.9&      7.850&      7.849 (  2.6)&     6.00    ( 1.36)&     7.66&    -1.67 &      -1.2\\
                        Al12HeLaf&   Al XII  &  6.9&      7.872&      7.869 (  4.5)&     4.71    ( 1.35)&     4.53&     0.18 &       0.1\\
                        Fe24w7.99&   Fe XXIV &  7.4&      7.991&      7.989 (  1.6)&     9.48    ( 1.31)&    13.82&    -4.34 &      -3.3\\
                        Fe24w8.23&   Fe XXIV &  7.4&      8.233&      8.283 (  4.2)&     3.24    ( 1.32)&     4.96&    -1.71 &      -1.3\\
                        Fe24w8.28&   Fe XXIV &  7.4&      8.285&      8.304 (  2.5)&     6.57    ( 1.74)&     1.85&     4.72 &       2.7\\
                        Fe23w8.30&   Fe XXIII&  7.2&      8.304&      8.318 (  6.5)&     8.12    ( 1.95)&     8.36&    -0.24 &      -0.1\\
                        Fe24w8.32&   Fe XXIV &  7.4&      8.316&      8.237 (  1.8)&     5.41    ( 1.30)&    10.09&    -4.68 &      -3.6\\
                          Mg12HLa&   Mg XII  &  7.2&      8.422&      8.420 (  0.3)&    98.76    ( 3.20)&   102.70&    -3.94 &      -1.2\\
                        Fe23w8.81&   Fe XXIII&  7.2&      8.815&      8.814 (  1.7)&     8.34    ( 1.47)&     8.38&    -0.04 &      -0.0\\
                        Fe22w8.97&   Fe XXII &  7.1&      8.975&      8.975 (  1.3)&     9.95    ( 1.64)&     7.57&     2.38 &       1.4\\
                        Mg11HeLar&   Mg XI   &  6.8&      9.169&      9.168 (  0.5)&    57.25    ( 3.01)&    59.28&    -2.03 &      -0.7\\
                        Mg11HeLai&   Mg XI   &  6.8&      9.230&      9.231 (  2.1)&    14.32    ( 2.06)&     8.99&     5.33 &       2.6\\
                        Mg11HeLaf&   Mg XI   &  6.8&      9.314&      9.313 (  0.7)&    28.47    ( 2.07)&    26.81&     1.66 &       0.8\\
                          Ne10HLg&   Ne X    &  7.0&      9.708&      9.710 (  1.2)&    45.64    ( 6.46)&    28.24&    17.40 &       2.7\\
                          Ne10HLb&   Ne X    &  7.0&     10.239&     10.238 (  0.4)&    85.51    ( 3.80)&    89.03&    -3.52 &      -0.9\\
                       Fe24w10.62&   Fe XXIV &  7.4&     10.619&     10.620 (  0.9)&    61.70    ( 4.27)&    65.97&    -4.27 &      -1.0\\
                       Fe24w10.66&   Fe XXIV &  7.4&     10.663&     10.661 (  1.9)&    34.56    ( 3.54)&    34.61&    -0.05 &      -0.0\\
                       Fe17w10.77&   Fe XVII &  6.8&     10.770&     10.768 (  2.5)&    10.41    ( 2.09)&     9.08&     1.32 &       0.6\\
                       Fe19w10.82&   Fe XIX  &  6.9&     10.816&     10.814 (  2.5)&    11.44    ( 2.16)&    11.96&    -0.52 &      -0.2\\
                       Fe23w10.98&   Fe XXIII&  7.2&     10.981&     10.983 (  1.2)&    44.19    ( 7.54)&    44.10&     0.09 &       0.0\\
                       Fe23w11.02&   Fe XXIII&  7.2&     11.019&     11.015 (  3.8)&    24.62    ( 7.90)&    28.89&    -4.27 &      -0.5\\
                       Fe24w11.03&   Fe XXIV &  7.4&     11.029&     11.031 (  7.8)&    54.16    ( 9.48)&    42.19&    11.97 &       1.3\\
                       Fe24w11.18&   Fe XXIV &  7.4&     11.176&     11.173 (  0.6)&    72.06    ( 4.36)&    76.20&    -4.14 &      -1.0\\
                       Fe18w11.33&   Fe XVIII&  6.8&     11.326&     11.330 (  0.9)&    18.26    ( 2.64)&    18.61&    -0.35 &      -0.1\\
                       Fe18w11.53&   Fe XVIII&  6.8&     11.527&     11.529 (  2.2)&    19.78    ( 3.11)&    12.78&     7.00 &       2.3\\
                          Ne9HeLb&   Ne IX   &  6.6&     11.544&     11.549 (  1.2)&    28.16    ( 3.37)&    20.24&     7.92 &       2.3\\
                       Fe23w11.74&   Fe XXIII&  7.2&     11.736&     11.738 (  1.1)&    91.55    ( 4.78)&    91.19&     0.36 &       0.1\\
                       Fe22w11.77&   Fe XXII &  7.1&     11.770&     11.771 (  1.1)&    74.99    ( 4.45)&    70.06&     4.93 &       1.1\\
      Fe17w12.12\tablenotemark{k}&   Fe XVII &  6.7&     12.124&     12.130 (  0.9)&    251.30  ( 33.48)&    50.17&   201.13 &       6.0\\
         Ne10HLa\tablenotemark{k}&   Ne X    &  6.9&     12.135&     12.139 (  3.8)&    383.89  ( 33.00)&   633.70&  -249.81 &      -7.6\\
                       Fe23w12.16&   Fe XXIII&  7.2&     12.161&     12.155 (  1.9)&     56.61 (  10.70)&    50.26&     6.35 &       0.6\\
                       Fe17w12.27&   Fe XVII &  6.7&     12.266&     12.261 (  3.4)&     27.94 (   9.27)&    45.09&   -17.16 &      -1.9\\
                       Fe21w12.28&   Fe XXI  &  7.1&     12.284&     12.285 (  0.6)&    137.49 (  15.15)&   135.90&     1.59 &       0.1\\
                       Fe20w12.58&   Fe XX   &  7.0&     12.576&     12.574 (  0.0)&     21.54 (   3.34)&    22.67&    -1.13 &      -0.3\\
                       Fe22w12.75&   Fe XXII &  7.1&     12.754&     12.751 (  1.9)&     23.90 (   3.63)&    25.34&    -1.44 &      -0.4\\
                         Ne9HeLar&   Ne IX   &  6.6&     13.447&     13.445 (  0.6)&    132.67 (  11.30)&   144.90&   -12.23 &      -1.1\\
                       Fe19w13.50&   Fe XIX  &  6.9&     13.497&     13.504 (  0.9)&     69.49 (   7.51)&    44.98&    24.50 &       3.3\\
                       Fe19w13.52&   Fe XIX  &  6.9&     13.518&     13.520 (  0.6)&     81.26 (   7.61)&    99.15&   -17.89 &      -2.4\\
                         Ne9HeLai&   Ne IX   &  6.6&     13.552&     13.554 (  1.9)&     36.59 (   4.81)&    19.55&    17.04 &       3.5\\
                         Ne9HeLaf&   Ne IX   &  6.6&     13.699&     13.699 (  0.6)&    110.30 (   7.00)&    65.60&    44.70 &       6.4\\
                       Fe18w14.26&   Fe XVIII&  6.8&     14.256&     14.205 (  0.3)&    119.10 (   8.24)&    40.82&    78.28 &       9.5\\
                       Fe20w14.27&   Fe XX   &  7.0&     14.267&     14.259 (  1.6)&     48.42 (   6.15)&    26.94&    21.48 &       3.5\\
                       Fe18w14.53&   Fe XVIII&  6.8&     14.534&     14.540 (  1.3)&     52.72 (   5.78)&    41.11&    11.61 &       2.0\\
                            O8HLd&   O  VIII &  6.7&     14.821&     14.820 (  0.0)&     21.59 (   4.95)&    17.90&     3.69 &       0.7\\
                       Fe17w15.01&   Fe XVII &  6.7&     15.014&     15.012 (  0.6)&    271.60 (  12.64)&   441.60&  -170.00 &     -13.4\\
                       Fe19w15.08&   Fe XIX  &  6.9&     15.079&     15.079 (  1.3)&     44.58 (   6.18)&    33.30&    11.28 &       1.8\\
           O8HLg\tablenotemark{l}&   O  VIII &  6.7&     15.176&     15.180 (  2.2)&     82.16 (   8.92)&    40.88&    41.28 &       4.6\\
                       Fe17w15.26&   Fe XVII &  6.7&     15.261&     15.261 (  1.2)&    116.33 (   9.19)&   124.40&    -8.07 &      -0.9\\
                       Fe18w15.62&   Fe XVIII&  6.8&     15.625&     15.625 (  1.9)&     42.85 (   6.16)&    55.81&   -12.96 &      -2.1\\
                       Fe18w15.82&   Fe XVIII&  6.8&     15.824&     15.826 (  2.2)&     32.92 (   5.94)&    33.99&    -1.06 &      -0.2\\
                       Fe18w15.87&   Fe XVIII&  6.8&     15.870&     15.869 (  1.9)&     33.99 (   6.10)&    18.03&    15.96 &       2.6\\
           O8HLb\tablenotemark{m}&   O  VIII &  6.7&     16.006&     16.005 (  0.9)&    130.71 (  26.12)&   127.10&     3.61 &       0.1\\
                       Fe18w16.07&   Fe XVIII&  6.8&     16.071&     16.072 (  0.9)&     99.86 (   9.51)&    77.00&    22.86 &       2.4\\
                       Fe19w16.11&   Fe XIX  &  6.9&     16.110&     16.106 (  3.4)&     25.85 (   6.84)&    43.53&   -17.68 &      -2.6\\
                       Fe18w16.16&   Fe XVIII&  6.8&     16.159&     16.168 (  3.7)&     21.99 (   6.20)&    31.16&    -9.17 &      -1.5\\
                       Fe17w16.78&   Fe XVII &  6.7&     16.780&     16.773 (  1.2)&    168.16 (  12.85)&   193.60&   -25.44 &      -2.0\\
                       Fe17w17.05&   Fe XVII &  6.7&     17.051&     17.049 (  0.6)&    236.37 (  14.89)&   232.30&     4.07 &       0.3\\
                       Fe17w17.10&   Fe XVII &  6.7&     17.096&     17.094 (  0.6)&    222.20 (  14.59)&   210.10&    12.10 &       0.8\\
                       Fe18w17.62&   Fe XVIII&  6.8&     17.623&     17.619 (  2.5)&     37.17 (   8.06)&    55.56&   -18.39 &      -2.3\\
                           O7HeLb&   O  VII  &  6.4&     18.627&     18.625 (  7.5)&     25.34 (   8.67)&    15.36&     9.98 &       1.2\\
                            O8HLa&   O  VIII &  6.7&     18.970&     18.968 (  2.0)&    810.96 (  87.37)&   858.80&   -47.84 &      -0.5\\
                          O7HeLar&   O  VII  &  6.3&     21.602&     21.601 (  1.9)&    125.43 (  22.71)&   128.90&    -3.47 &      -0.2\\
                          O7HeLai&   O  VII  &  6.3&     21.802&     21.804 (156.2)&     23.69 (  15.62)&    18.70&     5.00 &       0.3\\
                          O7HeLaf&   O  VII  &  6.3&     22.098&     22.095 ( 11.3)&     43.55 (  19.63)&    76.97&   -33.42 &      -1.7\\
                            N7HLa&   N  VII  &  6.5&     24.782&     24.780 (  3.1)&    132.88 (  25.49)&   137.20&    -4.32 &      -0.2\\
                       Fe18w93.92&   Fe XVIII&  6.8&     93.923&     93.920 (  3.0) &    661.91 ( 118.20)&   711.70&   -49.79 &      -0.4\\
                      Fe19w101.55&   Fe XIX  &  6.9&    101.550&    101.510 ( 40.0) &    259.18 (  78.54)&   235.00&    24.18 &       0.3\\
                      Fe18w103.94&   Fe XVIII&  6.8&    103.937&    103.940 ( -3.0) &    250.63 (  86.43)&   260.50&    -9.87 &      -0.1\\
                      Fe19w108.37&   Fe XIX  &  6.9&    108.370&    108.230 (140.0) &    796.49 ( 124.45)&   900.40&  -103.91 &      -0.8\\
                      Fe19w109.97&   Fe XIX  &  6.9&    109.970&    109.790 (180.0) &    217.01 (  74.83)&   120.20&    96.81 &       1.3\\
                      Fe20w110.63&   Fe XX   &  7.0&    110.630&    110.830 (-200.0)&    181.00 (  78.69)&    42.81&   138.19 &       1.8\\
                      Fe22w114.41&   Fe XXII &  7.1&    114.410&    114.430 (-20.0) &    339.81 (  94.39)&   358.90&   -19.09 &      -0.2\\
                      Fe20w118.66&   Fe XX   &  7.0&    118.660&    118.520 (140.0) &    284.33 (  98.05)&   528.10&  -243.77 &      -2.5\\
                      Fe19w120.00&   Fe XIX  &  6.9&    120.000&    119.810 (190.0) &    177.00 (  93.16)&   243.60&   -66.60 &      -0.7\\
                      Fe20w121.83&   Fe XX   &  7.0&    121.830&    121.750 ( 80.0) &    553.19 ( 134.93)&  1029.00&  -475.81 &      -3.5\\
                      Fe21w128.73&   Fe XXI  &  7.0&    128.730&    128.560 (170.0) &   1445.10 ( 190.13)&  1520.00&   -74.90 &      -0.4\\
                      Fe22w135.78&   Fe XXII &  7.1&    135.780&    135.590 (190.0) &   1168.80 ( 201.53)&   799.30&   369.50 &       1.8\\
                      Fe21w142.16&   Fe XXI  &  7.0&    142.215&    142.380 (-165.0)&    437.97 ( 199.07)&   139.70&   298.27 &       1.5\\
                       Fe9w171.07&   Fe IX   &  5.8&    171.073&    170.950 (123.0) &   1240.10 ( 539.18)&  1377.00&  -136.90 &      -0.3\\
                      Fe24w192.04&   Fe XXIV &  7.3&    192.017&    192.560 (-543.0)&   7453.50 ( 837.47)&  3120.00&  4333.50 &       5.2\\
                      Fe12w195.12&   Fe XII  &  6.1&    195.118&    195.920 (-802.0)&   1964.50 ( 577.78)&  1759.00&   205.50 &       0.4\\
                      Fe24w255.10&   Fe XXIV &  7.3&    255.090&    254.730 (360.0) &   3737.00 (1038.03)&  1655.00&  2082.00 &       2.0\\
                      Fe15w284.15&   Fe XV   &  6.4&    284.160&    284.110 ( 50.0) &   1945.40 ( 926.38)&  3024.00& -1078.60 &      -1.2\\
                      Fe16w335.41&   Fe XVI  &  6.5&    335.410&    334.210 (1200.0)&   8023.00 (1573.00)&  3892.00&  4131.00 &       2.6\\
                      Fe16w360.80&   Fe XVI  &  6.5&    360.761&    358.690 (2071.0)&   3941.00 (1115.90)&  2004.00&  1937.00 &       1.7\\
\enddata

\tablecomments{Values given in parentheses are the one standard
  deviation uncertainties on the preceding quantity. 
  Lines with wavelength greater than 30 \AA\ are EUVE data.}

\tablenotetext{a}{The mnemonic is a convenience for uniquely naming
  each feature.  It is comprised of the element and ion (in Arabic
  numerals) followed by a string indicating a wavelength and the
  wavelength (e.g., w16.78), or a code for the hydrogen-like (``H'') and
  helium-like ``He'' series, ``L'' for Lyman transition, one of ``a'',
  ``b'', ``g'', ``d'', or ``e'' for
  series lines $\alpha,\,\beta,\,\gamma,\,\delta,\,\epsilon$, and ``r'', ``i'',
  or ``f'' for resonance, intersystem, or forbidden lines. }

\tablenotetext{b}{Average logarithmic temperature [Kelvins] of
  formation, defined as the first moment of the emissivity distribution.}

\tablenotetext{c}{Theoretical wavelengths of identification (from APED), in
  \AA. If the line is a multiplet, we give the wavelength of the
  strongest component.}

\tablenotetext{d}{Measured wavelength, in \AA\  (uncertainty
  is in m\AA).}

\tablenotetext{e}{Emitted source line flux is $10^{-6}$ times the tabulated value in 
  $[\mathrm{phot\,cm\mtwo\,s\mone}]$.}

\tablenotetext{f}{Model line flux is $10^{-6}$ times the tabulated value in 
  $[\mathrm{phot\,cm\mtwo\,s\mone}]$.}

\tablenotetext{g}{Line flux residual, $\delta f = f_o - f_t$.}

\tablenotetext{h}{$\delta\chi = (f_o - f_t)/\sigma_o$.}

\tablenotetext{i}{\eli{S}{16} 3.991 \AA\ is blended with \eli{Ar}{17}
  3.994 \AA\ in about equal strengths for the model \dem.}

\tablenotetext{j}{\eli{Al}{13} 1.172 is blended with \eli{Fe}{24}
  7.169. We have adjusted the Al flux to account for a 25\%
  contribution of Fe for the assumed \dem.}

\tablenotetext{k}{\eli{Ne}{10} 12.132 \AA\ is blended with
  \eli{Fe}{17} 12.124 \AA; in the \heg\ spectrum, the lines are barely
  resolved, and expected to be about 10\% as strong.  Their sum has
  $\delta\chi=-0.3$.}

\tablenotetext{l}{\eli{O}{8} 15.18 \AA\ is blended with \eli{Fe}{19}
  in about equal strengths.}

\tablenotetext{m}{\eli{O}{8} 16.01 \AA\ is blended with \eli{Fe}{18};
  \eli{Fe}{18} 15.62 \AA\ has a similar theoretical emissivity
  distribution to \eli{Fe}{18} 16.01, so we have subtracted the
  latter's flux from the measured \eli{O}{8} to approximate the net
  \eli{O}{8} flux.}

\end{deluxetable}


\clearpage 
%
\begin{deluxetable}{rrcc}
\tablecaption{Elemental Abundances\label{tbl:abund}}
 \tablewidth{3.25in}
\tablehead{
  \colhead{Element} &
  \colhead{FIP\tablenotemark{a}} &
  \colhead{Corona\tablenotemark{b}} &
  \colhead{Photosphere\tablenotemark{c}} 
}
\startdata
  N &    14.53 & 1.2  (0.3)   &\\ 
  O &    13.62 & 0.6  (0.1)   &\\ 
 Ne &    21.56 & 1.6  (0.3)   &\\ 
 Mg &     7.65 & 0.7  (0.1)   & 1.0 (0.2)\\ 
 Al &     5.99 & 1.3  (0.3)   &\\ 
 Si &     8.15 & 0.6  (0.1)   & 1.3 (0.9)\\ 
 S  &    10.36 & 0.6  (0.1)   &\\ 
 Ar &    15.75 & 1.0  (0.2)   &\\ 
 Ca &     6.11 & 1.0  (0.3)   & 1.3 (0.4)\\ 
 Fe &     7.87 & 0.5  (0.1)   & 1.0 (0.2)\\ 
\enddata
\tablenotetext{a}{FIP is the first ionization potential in eV.}
\tablenotetext{b}{Coronal abundances are fractional relative to the Solar
 abundances of \citet{Anders:89}. Numbers
 in parentheses are one standard deviation uncertainties
 determined from Monte Carlo fitting.}
\tablenotetext{c}{Photospheric abundances from \citet{Gehren:99}.}
\end{deluxetable}


\newcommand{\CapFlux}{
  This is the energy flux density spectrum for the summed \heg\ and
  \meg\ first orders for all observations, totalling 97 ks. The
  spectrum has been smoothed by convolving it with a Gaussian kernel of
  dispersion 0.005\AA.  Significant lines have been marked.  The
  strength of the short-wavelength continuum ($<5$ \AA) and the number
  of high-excitation iron lines are indicative of the presence of very
  hot plasma up to about $100\,\mathrm{MK}$, while presence of
  \eli{N}{7}, \eli{O}{7}, \eli{O}{8}, and \eli{Ne}{9} indicates cooler
  plasma, near $3\,\mathrm{MK}$.}

\newcommand{\CapLc}{ We show the light curve over the six observations
  in several forms. The top panel shows the count rate against a
  binary system orbit number, with zero defined as a primary
  conjunction (HJD 2448501.1232).  Given the orbital period of 1.98318
  days, an orbit is 171 ks.  Both \meg\ and \heg\ orders $-3$ to $+3$
  (excluding the zero order, which is saturated) were summed into 0.5
  ks bins over the wavelength range of 1.7--25 \AA. Data are labeled
  by their \obs. The \euve\ light curve is plotted with small open
  squares (colored green in the electronic edition), and runs from
  about orbit 1.6 to 3.0, and is in 0.6 ks 
  bins. The \euve\ data's error bars, which have been omitted for
  clarity, are all less than 15\%. Note the small \euv\ flux
  enhancement during the flare in \obs\ 7.  The next panel shows the
  same data, but phase-folded, and only over the phase interval
  observed by \chan.  The dotted curve (solid orange in the electronic
  edition) is a simple
  occultation model for uniform disks of equal surface brightness to
  show where photospheric eclipses occur, arbitrarily scaled to
  exaggerate the eclipses.  The primary eclipse ($\mathrm{phase}=0.0$) is
  total, and the secondary is annular.  The \euve\ data are the sparse
  set of squares running along the bottom (colored green in the
  electronic edition).  The bottom pair
  of graphs show the light curves for narrow bands around \eli{Si}{14}
  (6.16--6.22 \AA), which is strongly modulated (upper pane), and
  \eli{Si}{13} (6.62--6.78 \AA), which is only weakly modulated
  (bottom pane).  Each has had a nearby continuum band rate scaled and
  subtracted.  Different orbits at the same phase are distinguished by
  the error-bar line style (solid blue in the electronic edition).  }

\newcommand{\CapMod}{We computed the modulation of features from their
  count-rate, defined as the $(max-min)/(max+min)$.  Lines are labeled
  by element and ion (in Arabic notation). These are plotted against
  their temperature of maximum emissivity.  The hotter plasma is more
  strongly modulated, which is consistent with variations being due to
  flares, rather than a change in volume which would affect all lines
  equally.} 

\newcommand{\CapCounts}{These are the \meg\ and \heg\ counts spectra
  and the folded \dem\ and abundance model for the summed $+1$ and
  $-1$ orders.  Both data and model have been smoothed by a Gaussian
  convolution with dispersions of 0.005 \AA\ (\meg) and 0.0025 \AA\ 
  (\heg).  The model is plotted with a dashed line (solid red in the
  electronic edition).  The \meg\ 
  spectrum is usually the upper trace, except near \eli{Fe}{25}, 1.85
  \AA, where the \heg\ has more effective area.  \heg\ has about twice
  the resolution of \meg.  Some model lines expected in each region
  have been labeled.  Some regions of the spectrum are well matched by
  the model.  In others, large mismatches are obvious. The spectra are
  shown in four parts, a-d, each of which has four panels covering
  about 1.5 \AA. Part a spans 1.7--7.7 \AA, part b 7.7--13.7 \AA, part
  c 13.7--19.7, and part d 19.7--25.7 \AA.} 

  \newcommand{\CapCountsB}{See~\ref{fig:counts}}
  \newcommand{\CapCountsC}{See~\ref{fig:counts} }
  \newcommand{\CapCountsD}{See~\ref{fig:counts} }

  \newcommand{\CapDEM}{We show the differential emission measure
    integrated in bins of size $\Delta\log T = 0.05 $ for the DEM fit
    to combined observations.  The solid line is the mean of the 100
    Monte-Carlo fits, and the dashed lines are the one standard
    deviation boundaries.  The tail above about $\log T=7.6$ was
    manually adjusted (within the one sigma limits) so that the
    synthetic spectrum's short wavelength continuum better matched the
    data, since this wasn't well constrained by the fit.  The peak at
    6.2 is constrained by contemporaneous \euve\ data.  The range
    including the hot peak at 7.3 and hotter tail is strongly
    modulated by flares.  The integral over the plotted temperature
    range yields a volume emission measure of
    $1.2\times10^{54}\,\mathrm{cm\mthree}$.  }
  
  \newcommand{\CapAbund}{The abundances relative to Solar are plotted
    against their first ionization potential.  Results from the
    current work are marked with squares, and the uncertainties are
    one standard deviation from the Monte Carlo fit.  Results from
    low-resolution X-ray determinations are also plotted (diamonds and
    asterisks, with dotted error-bars) as well as the AR Lac
    photospheric values (triangles) and scaled Solar coronal
    abundances (dashed line); the symbol legend indicates the first
    author.}

  \newcommand{\CapLineProf}{In this figure we compare the normalized
    flux profiles at quadrature and conjunction phases.  Normalization
    is done within the wavelength band by subtracting the minimum and
    then dividing by the maximum.  The solid line profile is from the
    conjunction phases when the stars' line-of-sight velocity
    difference is zero, the dashed quadrature, and the lowest dash-dot
    line is the difference.  The left panel is from \meg\, and the
    right is \heg\ (at the same wavelength, \heg\ has twice the
    resolution of the \meg). Both broadening and shifts are apparent.}

  \newcommand{\oldCapLineProf}{In this figure we compare the normalized
    flux profiles at quadrature and conjunction phases.  Normalization
    is done within the wavelength band by subtracting the minimum and
    then dividing by the maximum.  The solid line profile is from the
    conjunction phases when the stars' line-of-sight velocity
    difference is zero, the dashed quadrature, and the lowest dash-dot
    line is the difference.  The upper two panels are from \meg\, and
    the lower two are \heg\ (at the same wavelength, \heg\ has twice
    the resolution of the \meg). Both broadening and shifts are apparent.  }
  
  \newcommand{\CapLchi}{In this figure we plot the line flux residuals
    normalized by their measurement uncertainty (upper graph) and the
    ratio of measured line flux to model flux (bottom graph).  The
    abscissa is the first moment of the line's emissivity
    distribution, rather than the temperature of maximum emissivity.
    This is a conservative bias for the hydrogen-like ions, since they
    have a long tail on the high-temperature side they move to
    slightly higher values than their peak.  In the both plots, each
    ion is denoted by the first character in its element abbreviation
    with some use of lower-case for duplicate characters (see the
    legend).  While many features lie within $\pm1\sigma$ of zero,
    many lie outside two and 3$\sigma$, in excess of what would be
    expected statistically for Gaussian measurement errors alone.
    Hence, there is a deficiency in the \dem\ and abundance model,
    errors in the atomic data, mis-identified and blended lines, or
    likely, some combination of these.  }

\clearpage 

\figcaption[Espec.ps]{\CapFlux}
\figcaption[Lc_a.ps]{\CapLc}
\figcaption[Lc_Mod.ps]{\CapMod}
\figcaption[Counts_Spec_A_gs.ps]{\CapCounts}

\begin{figure}
  \figurenum{4b}
\epsscale{0.800}
\plotone{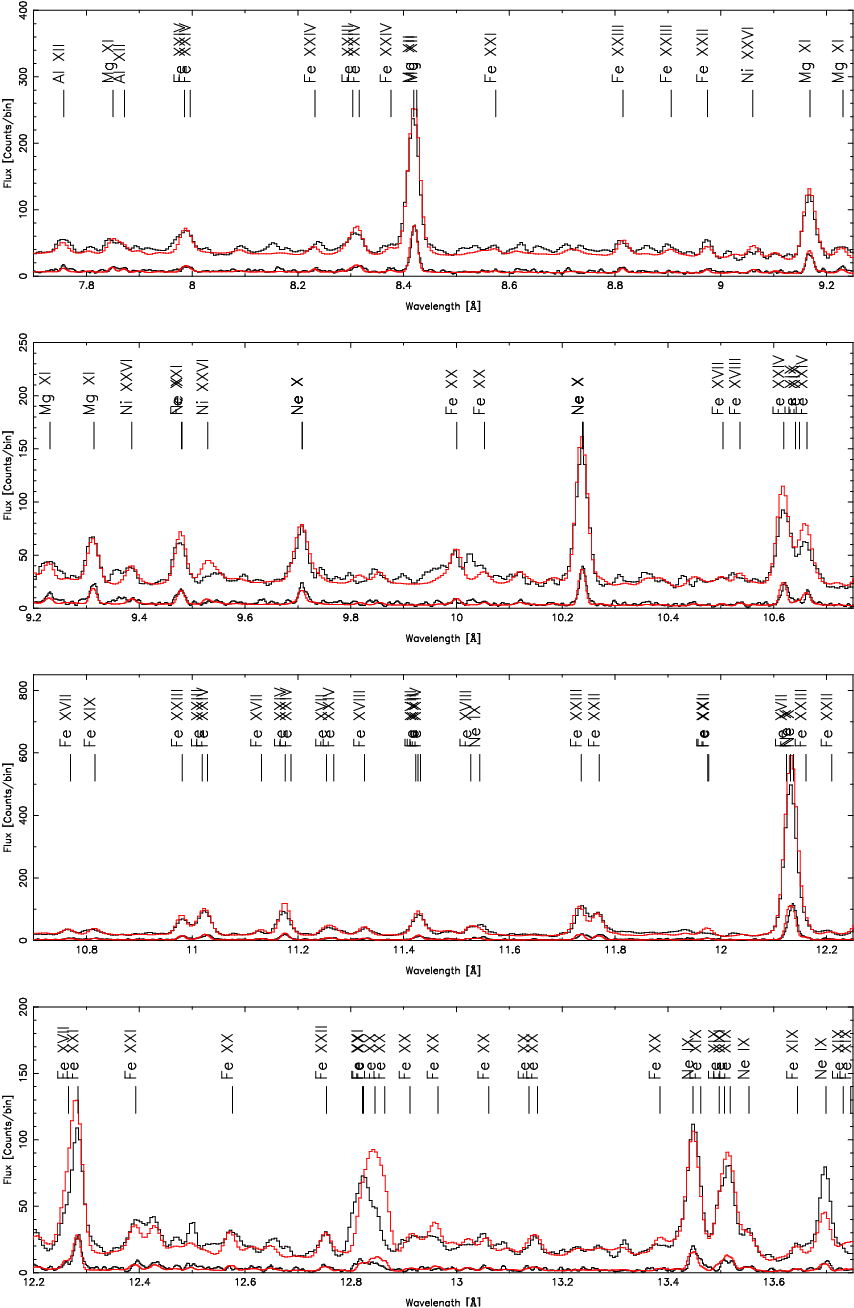}               
\caption{}
\label{fig:countsB}
\end{figure}

\figurenum{\ref{fig:countsB}}\figcaption[]{\CapCountsB}

\begin{figure}
  \figurenum{4c}
\epsscale{0.800}
\plotone{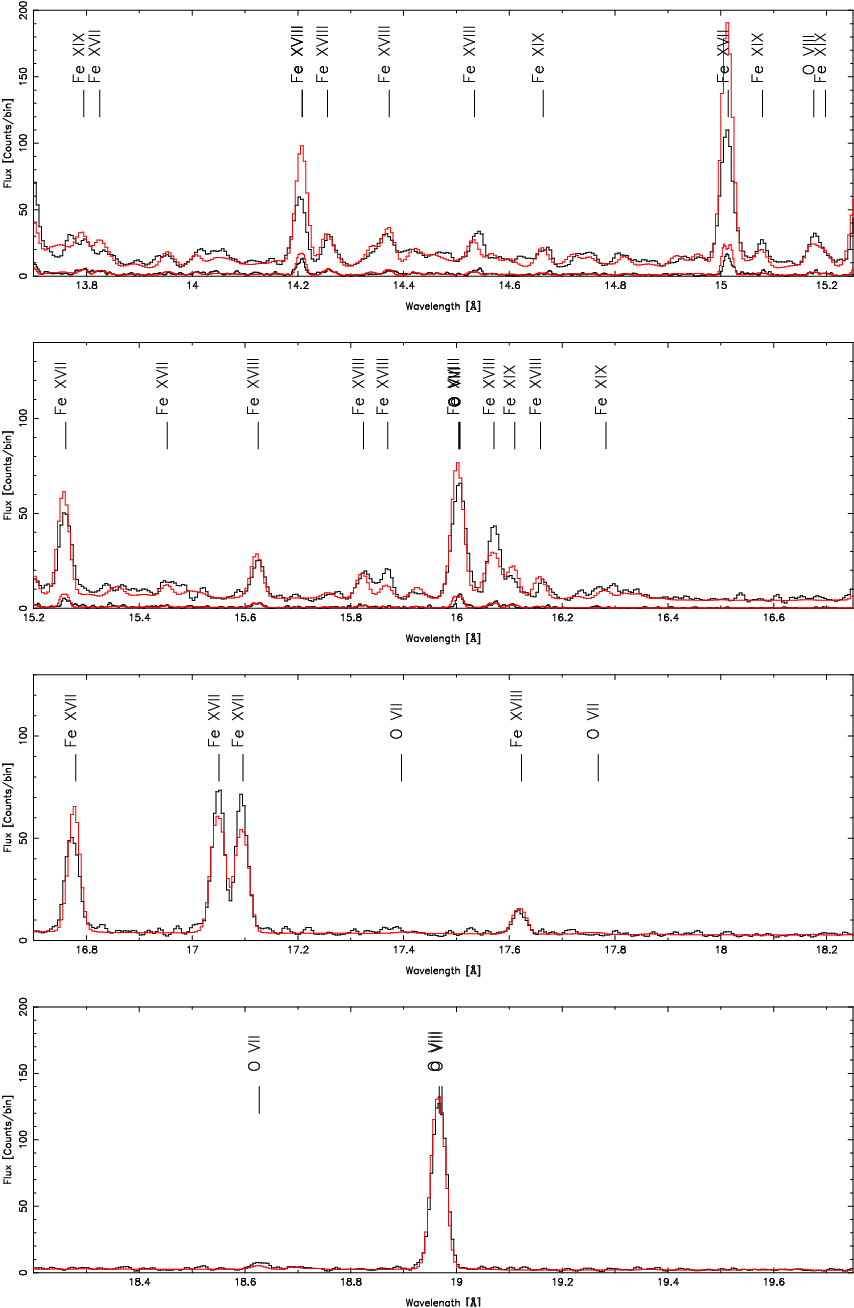}            
\caption{}
\label{fig:countsC}
\end{figure}

\begin{figure}
  \figurenum{4d}
\epsscale{0.800}
\plotone{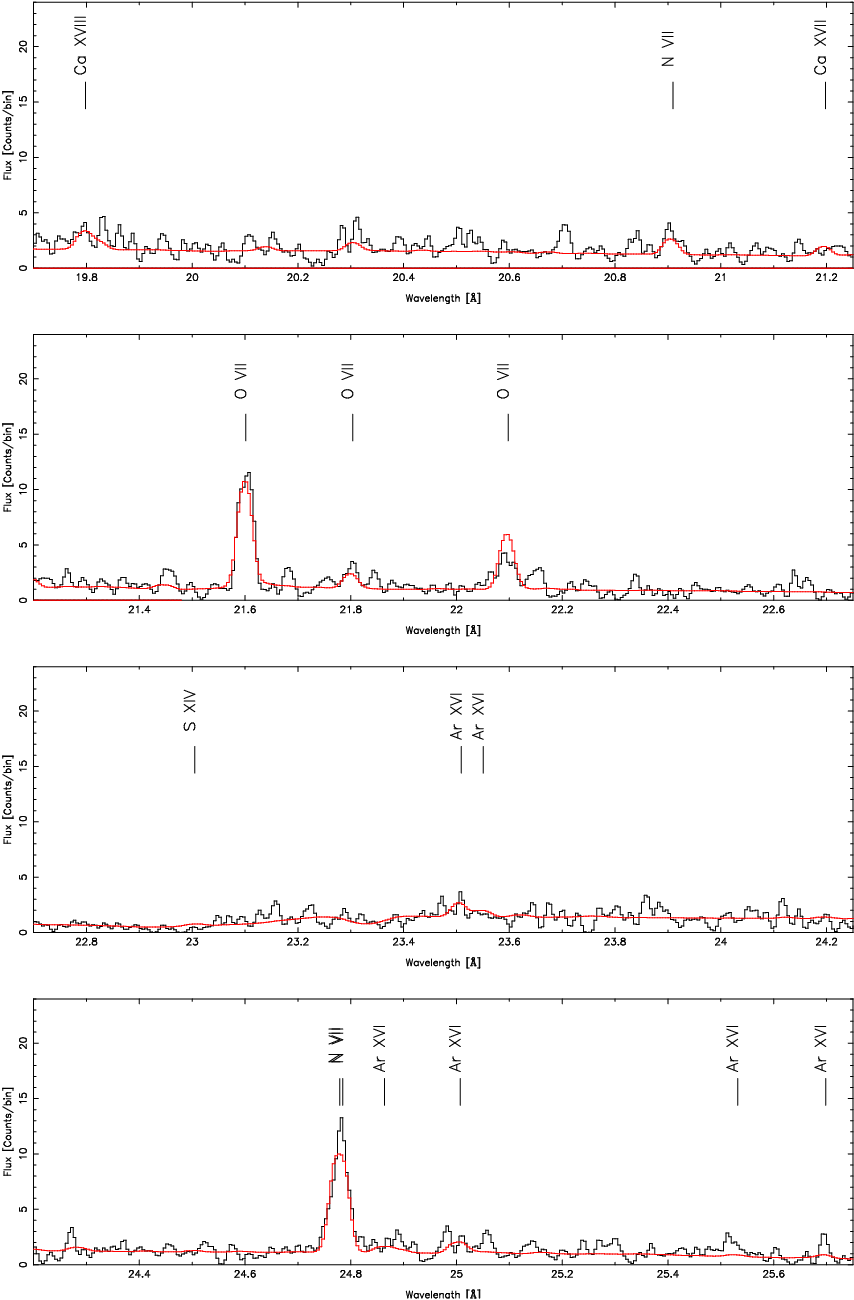}           
\caption{}
\label{fig:countsD}
\end{figure}

\begin{figure}
  \figurenum{6}
\plotone{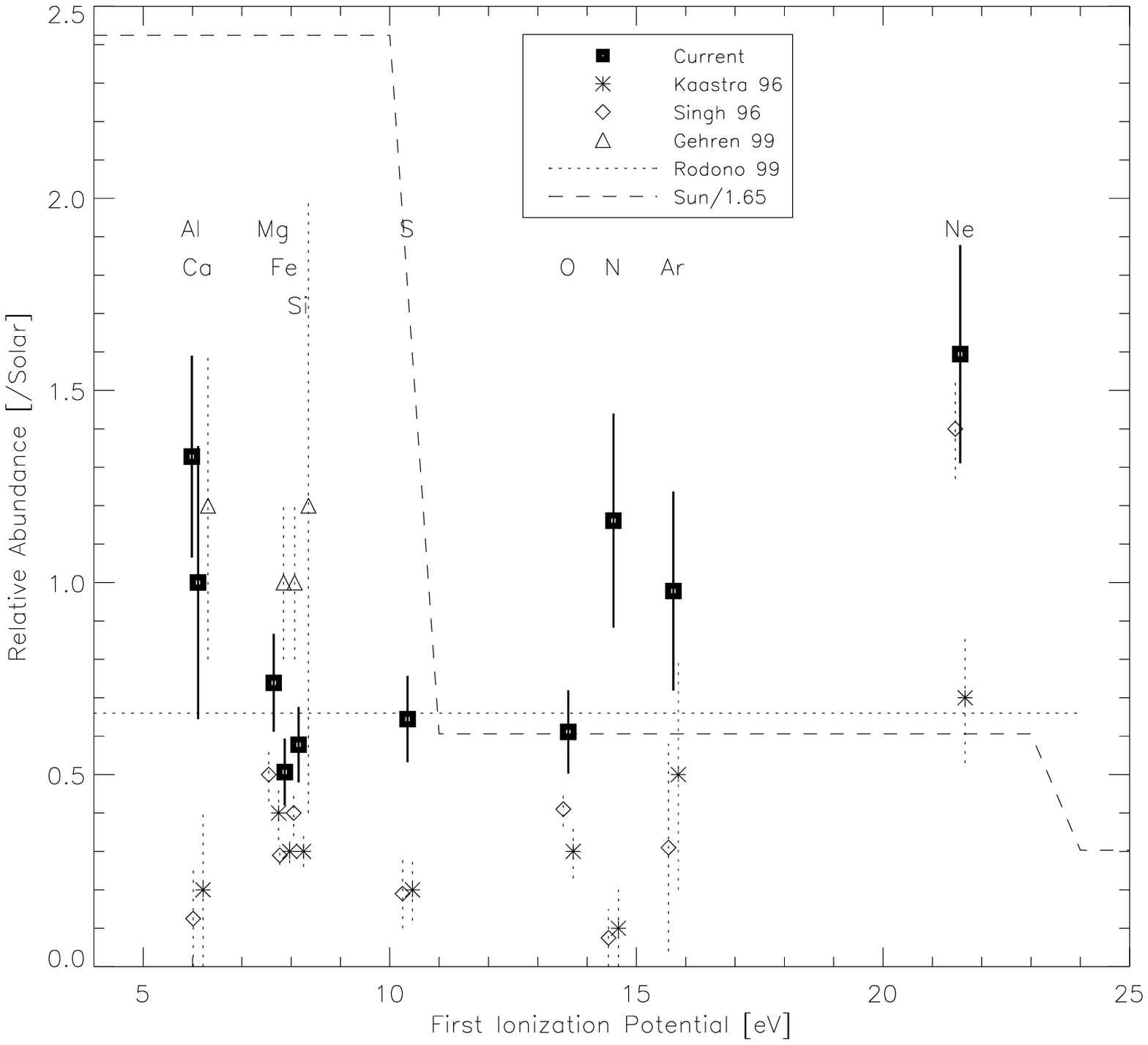}          
\caption{\CapAbund}
\label{fig:abund}
\end{figure}

\begin{figure}
  \figurenum{8}
\epsscale{0.700}
\plotone{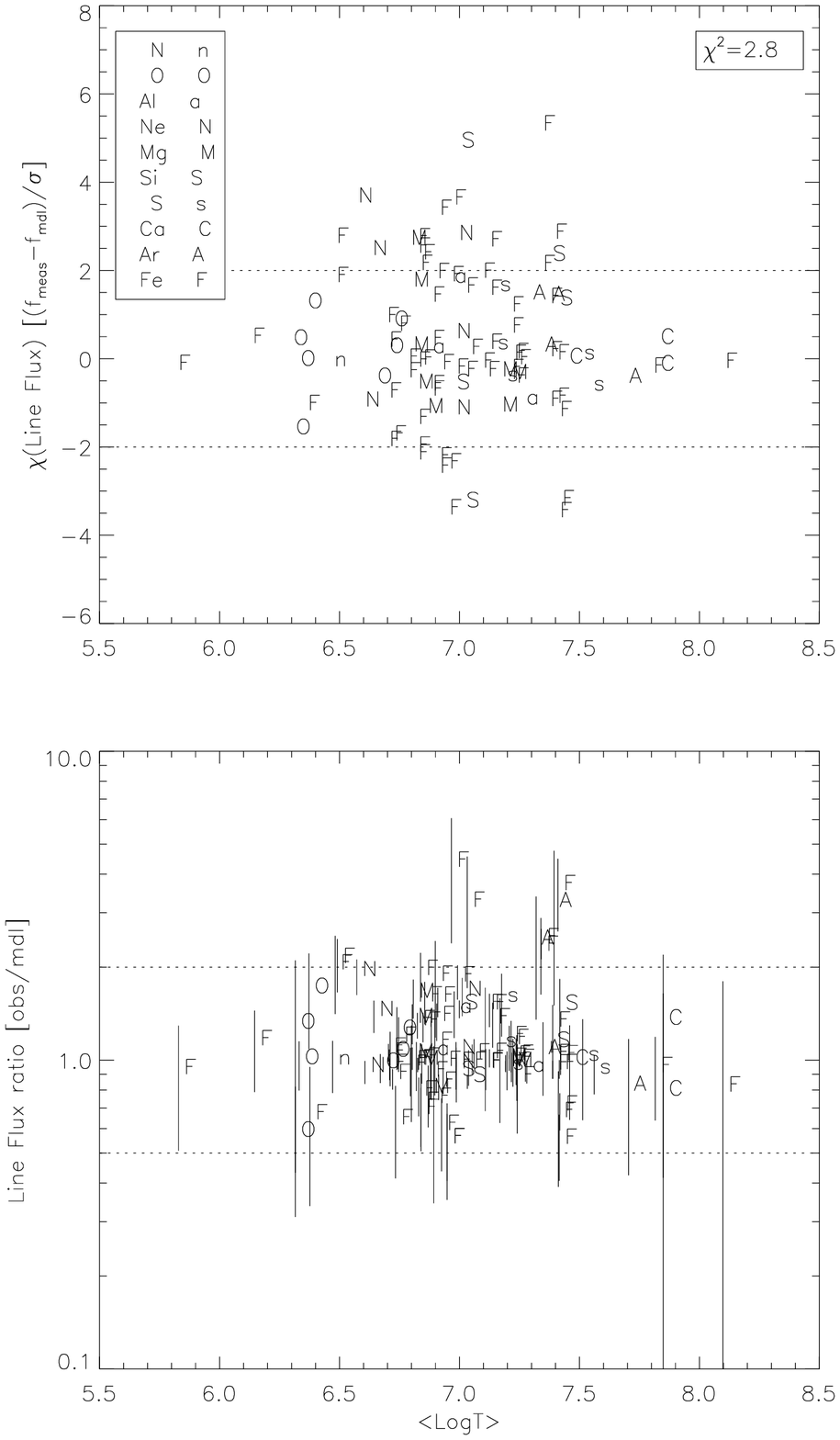}           
\caption{\CapLchi}
\label{fig:chiflux}.
\end{figure}

\figurenum{\ref{fig:countsC}}\figcaption[]{\CapCountsC}
\figurenum{\ref{fig:countsD}}\figcaption[]{\CapCountsD}
\figurenum{\ref{fig:dem}}\figcaption[DEM.ps]{\CapDEM}
\figurenum{\ref{fig:abund}}\figcaption[Abund.ps]{\CapAbund}
\figurenum{\ref{fig:broadlines}}\figcaption[Broadlines.ps]{\CapLineProf}
\figurenum{\ref{fig:chiflux}}\figcaption[Chi_Flux.ps]{\CapLchi}

\clearpage
\onecolumn
%
\begin{figure}
   \figurenum{1}
\plotone{f1.eps}  
   \caption{\CapFlux}
   \label{fig:flux}
 \end{figure}

\clearpage
%

\clearpage
\begin{figure}
  \figurenum{3}
\plotone{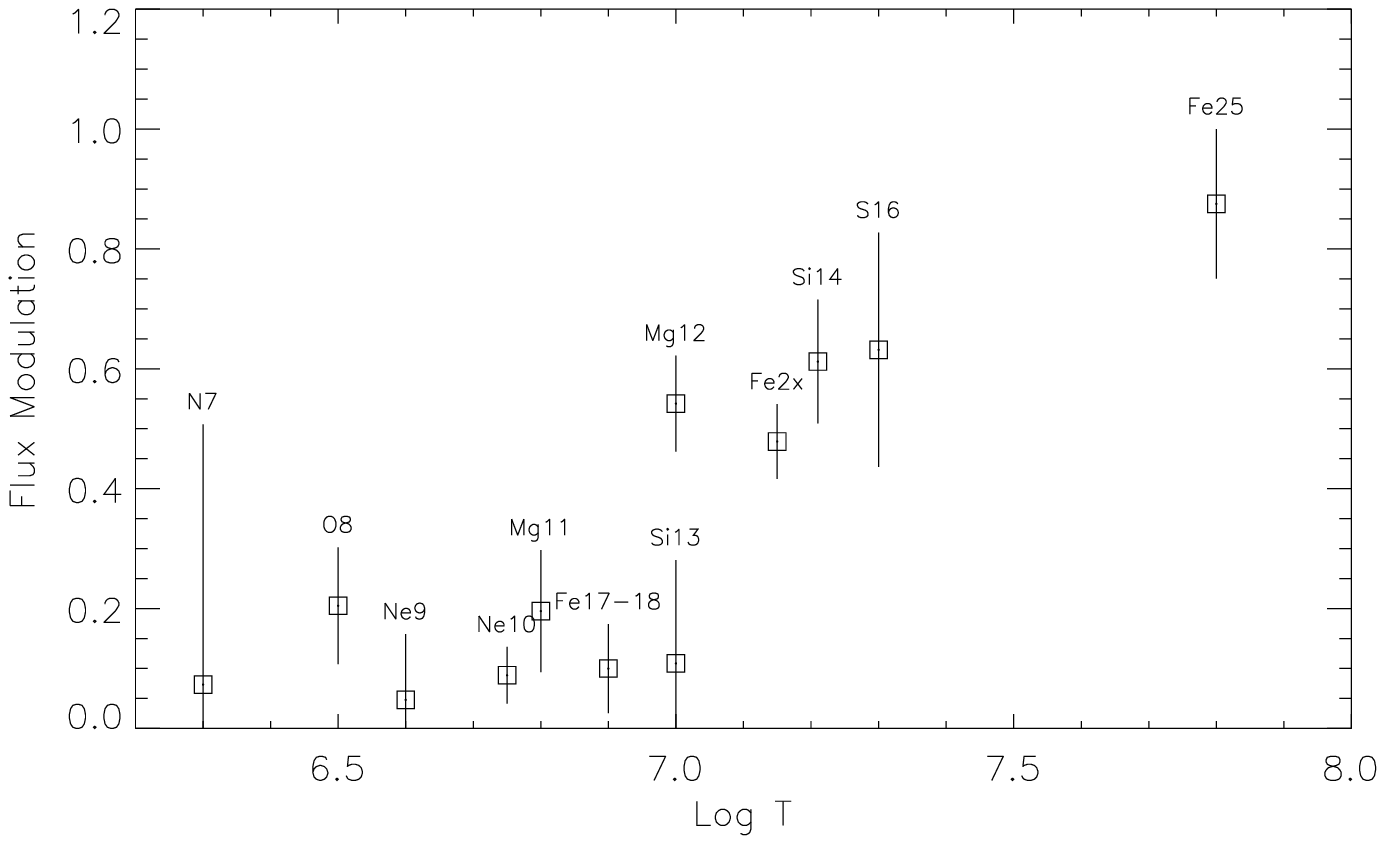} 
\caption{\CapMod}
\label{fig:lcmod}
\end{figure}

\clearpage
%
\begin{figure}
  \figurenum{4a}
\epsscale{0.800}
\plotone{f4a_c.eps}        
\caption{}
\label{fig:countsA}
\figurenum{4}\label{fig:counts}
\end{figure}

\clearpage

\clearpage

\clearpage

\clearpage
%
\begin{figure}
  \figurenum{5}
\plotone{f5.eps}          
\caption{\CapDEM}
\label{fig:dem}
\end{figure}

\clearpage
%

\clearpage
%
\begin{figure}
  \figurenum{7}
\plotone{f7.eps}            
\caption{\CapLineProf}
\label{fig:broadlines}
\end{figure}

\clearpage
%

\end{document}